\documentclass[12pt]{article}
\usepackage{amsmath}
\usepackage{graphicx}
\usepackage[margin=1in]{geometry}
\usepackage{natbib}
\usepackage{enumitem}
\usepackage{amssymb}
\usepackage{clipboard}
\usepackage{float}
\usepackage{url}
\usepackage{hyperref}
\usepackage{xcolor}
\usepackage{caption}
\usepackage{subcaption}
\usepackage{bibunits}

\newcommand{\blind}{0}
\newcommand{\iid}{\overset{iid}{\sim}}

\newcommand{\red}{\textcolor{black}}
\newcommand{\change}{\textcolor{black}}

\DeclareMathOperator*{\argmin}{arg\,min}

\begin{document}


\allowdisplaybreaks

\def\spacingset#1{\renewcommand{\baselinestretch}%
{#1}\small\normalsize} \spacingset{1}


\if0\blind
{
  \title{\bf The Impact of Stocks on Correlations between Crop Yields and Prices and on Revenue Insurance Premiums using Semiparametric Quantile Regression}
  \author{Matthew Stuart\\
    Department of Mathematics and Statistics, \\
    Center for Data Science and Consulting, Loyola University, Chicago \\
    1032 W. Sheridan Road \\ Chicago, IL 60660 \\
    \href{mailto:mstuart1@luc.edu}{mstuart1@luc.edu}
    \and
    Cindy Yu \\
    Department of Statistics, Iowa State University \\
    2431 Osborn Drive \\ Ames, IA 50011 \\
    \href{mailto:cindyyu@iastate.edu}{cindyyu@iastate.edu}
    \and
    David A. Hennessy \\
    Department of Economics, Iowa State University \\
    518 Farm House Ln. \\Ames, IA 50011 \\
    \href{mailto:hennessy@iastate.edu}{hennessy@iastate.edu}}
    \date{}
  \maketitle
} \fi

\if1\blind
{
  \title{\bf The Impact of Stocks on Correlations between Crop Yields and Prices and on Revenue Insurance Premiums using Semiparametric Quantile Regression}
    \date{}
  \maketitle
} \fi

\thispagestyle{empty}

\noindent {\it Abstract:}%

Crop yields and harvest prices are often considered to be negatively correlated, thus acting as a natural risk management hedge through stabilizing revenues. Storage theory gives reason to believe that the correlation is an increasing function of stocks carried over from previous years. Stock-conditioned second moments 
have implications for price movements during shortages and for hedging needs, while spatially varying yield-price correlation structures have implications for who benefits from commodity support policies. 
In this paper, we use semi-parametric quantile regression (SQR) with penalized B-splines to estimate a stock-conditioned joint distribution of yield and price. The method 
enables sampling from the \change{empirical} joint distribution using SQR. Then it is applied to approximate the stock-conditioned correlation 
for both corn and soybeans in the United States. For both crops, Cornbelt core regions have more negative correlations than do peripheral regions. We find strong evidence that correlation becomes less negative as stocks increase \change{and also upon moving north. We suggest three channels through which stocks can predict revenue; through price level, price variability and price-yield correlation. The first two channels are currently addressed in premium rate-setting procedures. The third is not and we provide yield autocorrelation evidence to suggest that this could be a concern. \red{We conduct a rating game to evaluate the validity of our methodology for assessing premium rates compared to the current methods used by the RMA.} Our results indicate that accounting for this third channel does not materially alter premium rates}. \\

\noindent {\it Keywords:}%

Core and periphery, Crop insurance, Density function estimation, Implied volatility, Storage theory, Tail Dependence. \\




\clearpage
\setcounter{page}{1}
\spacingset{2} 

\section{Introduction}\label{Introduction}

Storage infrastructure and the associated commodity stocks serve a critical role in ensuring resilience to commodity market shocks. Resilience in markets is reflected by the degree of price insensitivity to a shock, most typically a current-year yield response to underlying weather, disease or related production conditions. Price-yield correlation, being a normalized measure of how sensitive price is to a production shock, can measure both resilience and revenue stability for growers. \Copy{R1C1}{\change{Stocks carried into the current year can be drawn down to supplement consumption in an adverse yield event. How price-yield correlation changes with stocks is one component of how larger stocks contribute to manage market shocks. The literature on commodity price volatility and storage is large and the theoretical models are compelling. However, empirical analysis is often at variance with the models \citep{DeatonLaroque,DeatonLaroque1996,Wright2011}. As an empirical matter, little is known about how commodity stocks affect a commodity’s price-yield correlation that is central to stabilizing revenue and buffering price movements against a production shock.}} The intents of this paper are to accurately estimate the price-yield joint distribution when conditioned on stocks using semiparametric quantile regression methods, and then provide evidence on how stocks affect \red{price-yield correlation,} price variability, producer revenue variability, 
and revenue insurance premium calculations.

The matter is important for several reasons where perhaps the most important is in understanding the source of large and dramatic commodity price spikes. Price spikes often arise when stocks are low and are more frequent for commodities, such as electricity and fresh produce, that are difficult or impossible to store. For consumers, price variability reduces economic welfare under certain intuitive conditions \citep{Turnovsky}, while for risk averse producers who cannot adapt decisions to variable prices, the impact is detrimental \citep{Oi,BellemareBarrett}. A large, nuanced and as yet inconclusive literature has also investigated whether such price spikes act to cause political instability \citep{blair_christensen_rudkin_2021,FoodandConflict}.

A second motivation for inquiry into the issue regards our collective understanding of regional comparative advantage. Core-periphery theory emphasizes the advantages and disadvantages that a core region in an economic activity has relative to less central areas. Agglomeration economies, due to lower transportation, search, learning and other benefits associated with spatially concentrated economic activity, are often among the most prominent advantages possessed by the core \citep{Puga}. We follow others in documenting a natural hedge effect \citep{Finger2012} whereby revenue is more stable in core crop production regions than in the periphery \citep{RamsayGodwinGhosh}. For example, low price and low yield are less likely to occur simultaneously in a core region than in a region more distant from the center of production. In consequence, production in the core region is comparatively less financially risky. In this light the availability of opportunities to insure against revenue shortfalls can be seen as a counterweight to this advantage. We go further in showing that the regional comparative advantage provided by the natural hedge is likely to be most important whenever stocks carried into the year are low. 

\Copy{R1C2}{\change{An additional stimulus for this study is more directly practical, having to do with modeling assumptions when pricing revenue insurance (RI), which was first offered in the mid-1990s and quickly became the most popular form of crop insurance offered in the United States. As insurance is intended to indemnify in extreme events, modeling assumptions that apply on average but do not apply in tail events can prove to be problematic, \citep[see][among others]{Zimmer2012}. Biased probabilities of extreme outcomes may lead to biased premiums because payoffs are largest in extreme outcomes. Depending on the direction of any bias, the products may attract unduly high-risk business or lead to low participation. If high crop insurance participation is a government policy goal, as is true in many parts of the world, then that government may be required to provide larger subsidies to secure the desired participation level. Stated differently, removing any biases should allow for high participation rates but with lower subsidy transfer levels where premium subsidies for the major crops grown in the United States in 2023 were at 60\% or more.}

\change{A large literature has emerged to address the modeling of extreme events in crop insurance markets. \cite{Goodwin2001} has shown that county-level yield correlations vary spatially and strengthen during years of widespread drought. \cite{Andersonetal2013} note that certain inflexibilities in standard approaches to modeling revenue insurance correlation structures can be overcome by applying a method due to \cite{Phoon2004}. \cite{Duetal} use copula modeling approaches to study yield-yield dependence structures across nearby land units. \cite{Parketal2019} provide a modeling procedure that emphasizes the left tail of yield distributions, which is of highest interest in crop insurance. 
\cite{AFordRamsey} proposes a quantile-based approach that is flexible in how distribution conditioners, such as technology and weather, affect distribution tails. Our focus is more specific than any of the above in that we have a theory-grounded concern, namely that stocks (storage) alter the joint price-yield distribution in a definite way that may determine crop insurance premiums.}}

\Copy{R3SC1}{Accurately estimating the stock-conditioned joint price-yield distribution is crucial for calculating price-yield correlation and revenue insurance premiums\footnote{In their report to the USDA RMA, \cite{Goodwinetal2014} noted that price-yield relationships may not be constant over space or time and may vary with world stocks. That report recommended the RMA to be attentive to price-yield correlation modeling assumptions. By considering spatial patterns in price-yield correlations for corn and soybeans in the United States, \cite{RamsayGodwinGhosh} have filled in some of this identified gap in knowledge.}. Harvest price distributions are generally right-skewed and harvest yields are generally left-skewed \citep[see][among others]{HennessySkewness}. We do not want to misspecify a parametric model of the joint distribution, and subsequently, misidentify the correlation. In addition, a purely non-parametric approach will produce estimates that are heavily influenced by the sample size. To overcome these limitations, we propose a semiparametric quantile regression (SQR) approach that enables the sampling of valid draws from the target joint price-yield distribution, conditioned on stocks.
These draws can then be used to estimate the moments of various functions involving price and yield through Monte Carlo averaging. Examples of such moments include the stock-conditional correlation between price and yield, as \change{well} as stock-conditional crop premium rates (i.e., the mean of indemnity payout which depends on price and yield).
}

\Copy{R1C5}{
\change{
Our SQR method uses B-splines to estimate the $\tau$-th quantile function for different $\tau$ values ranging from 0 to 1, allowing for flexibility in identifying potential non-linear relationships \red{with heteroscedasticity, which are not easily captured by standard normal error models.} Outliers in the sampled draws can be easily avoided by controlling the simulated $\tau$ values to ensure that they are not too close to 0 or 1.
}
}

\Copy{R2C9}{We are not the first to use quantile regression in econometrics \citep{Koenker2000,KoenkerHallock}. \cite{HousePrices} \red{use} quantile regression techniques with spatial econometric modeling for hedonic pricing for housing valuation. \change{\cite{Kim2007} \red{proposes} a method for conditional quantiles using polynomial splines with varying coefficients, and performs tests for these estimates against linear quantile regression estimates. \cite{QAR} use quantile autoregression techniques to estimate price volatility based on stocks. \cite{FordRamseyBayesian} used a Bayesian quantile regression methodology to estimate univariate yield densities.} Our paper is distinctive from those papers in terms of objective and estimation method:  (1) We propose to simulate valid draws from a 2-dimensional joint density of crop yields and prices, while previous studies have focused only on estimating the univariate density of crop yields \change{or prices}, and (2) we propose to estimate the quantile regression functions semiparametrically using penalized B-splines, while others have used a strictly non-parametric approach \citep{AFordRamsey,Goodwin1998, Barnwal,KerandCoble}.}

\change{The remainder of the paper is organized as follows. Section \ref{Framework} discusses the conceptual framework of the theory of storage, and its implication for indemnities.  Section \ref{Methodology} outlines in detail the procedures for SQR, the construction of the B-spline bases, and the sampling procedure from the joint distribution function. Section \ref{Empirical} employs our method empirically on the corn crop. Section \ref{Conclusion} concludes with some final remarks.}

\section{Conceptual Framework} \label{Framework}

\subsection{Theory of Storage}

\begin{figure}[h!]
    \centering
    \includegraphics[scale=0.4]{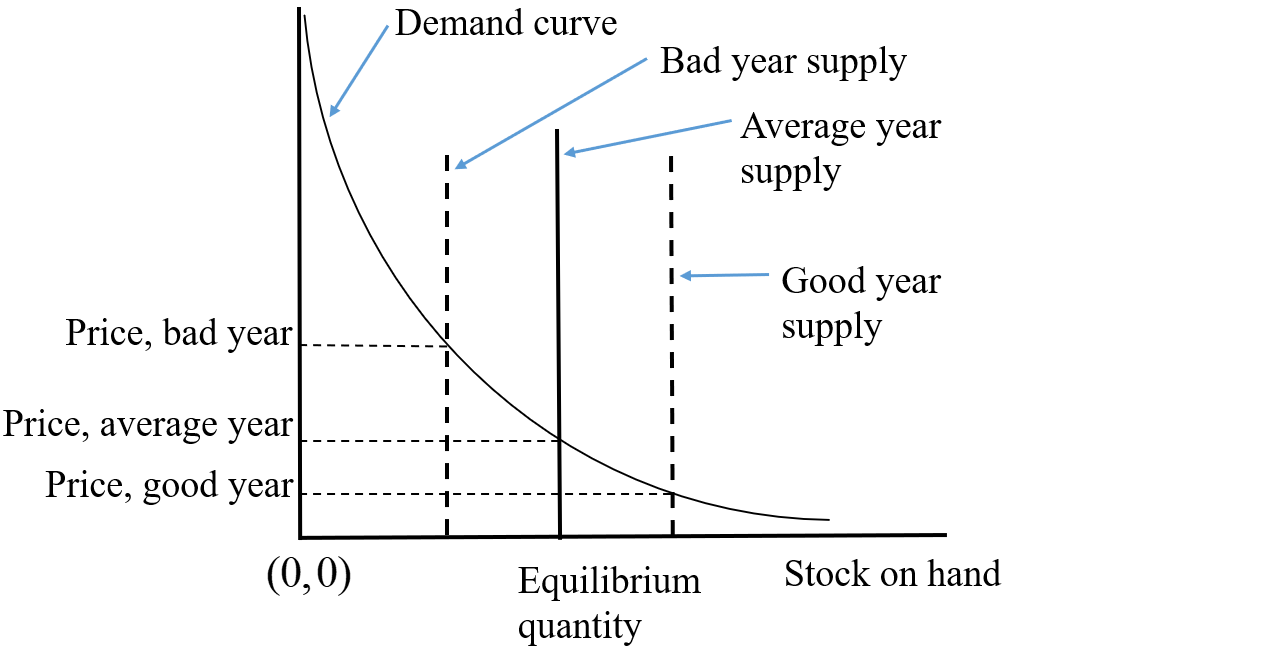}
    \caption{Demand curve and equilibrium price for a crop of interest when stocks are not incorporated in the demand curve.}
    \label{Figure1}
\end{figure}

\Copy{R1C9}{\change{The theory behind the role of stocks in determining second moments of the price yield distribution can be summarized in two graphics, the second being an adaptation of a well-known figure in \cite{DeatonLaroque}.}} For a given acreage allocated to a crop but random yield from these acres, Figure \ref{Figure1} visualizes the traditional supply and demand relationship absent stocks. The horizontal axis reflects current-year production only. Demand is a stable function of commodity available with finite negative slope. 

\Copy{R1C7}{\change{To understand the supply side characterization, the farmer’s management decisions may be partitioned into those that occur at or before planting and those that occur after planting. Fertilizer side-dressing, pest management and some other choices do occur after planting. Such decisions may be affected by the pertaining price level where in extreme cases a poor crop will not be harvested when prices are low. However, farm production is determined largely by either (i) acreage, seed and fertilizer input decisions made at or before planting, or (ii) weather and disease conditions that are often entirely exogenous to production. Indeed, much of the variation in post-planting input choices is in response to exogenous weather and disease events that may require a pest management intervention or preclude entry onto the land for scheduled events.}} For these reasons, the short-run supply function is held to be vertical, or unresponsive to price, and stochastic where this randomness is characterized by good-year and bad-year yield outcomes either side of the vertical average year supply curve. 

Visible in Figure \ref{Figure1} is what might be termed “the farmer’s curse” whereby a good harvest, presuming that it is good for all producers, leads to low prices so that revenue might decline in good years. If the demand curve is very inelastic, and so the correlation between a production shock and price is strongly negative, then a positive yield shock will decrease revenue. In fact this phenomenon is an extreme outcome from one of two manifestations of the natural hedge where the farmer must also recognize the associated event that prices rise when supplies are limited.

\begin{figure}[h!]
    \centering
    \includegraphics[scale=0.4]{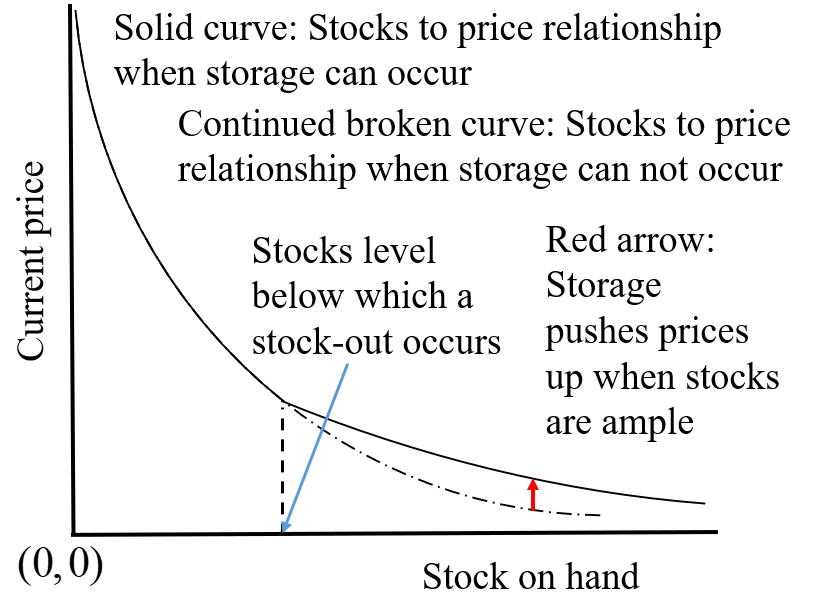}
    \caption{Demand curve and equilibrium price for a crop of interest with the inclusion of stocks in the demand curve.}
    \label{Figure2}
\end{figure}

Including storage fundamentally changes the price quantity relationship \citep{StuttgenBoatwright}. \Copy{R1C8}{\change{With storage then good-year product that would obtain a low price were it sold in that year can be stored in the reasonable expectation that price will be higher the following year.}} Then the stored product purchased at a low price can be sold at a price high enough to compensate for storage costs. The activity is inherently risky, however, because storage costs are certain but future harvests are random. A storer will incur a loss in the event of several good-year harvests. Figure \ref{Figure2} visualizes how \cite{DeatonLaroque} characterize the availability of stored stocks for consumption and how it 
would relate to current market price were decision makers acting rationally. When compared with Figure \ref{Figure1}, storage is allowed in Figure \ref{Figure2} so that the horizontal axis includes both (i) current-year quantity and (ii) stocks carried over from prior years. The downward sloping price-to-stocks curve in Figure \ref{Figure2} is a demand curve, but does not necessarily represent demand for current year consumption. Professional storers are assumed to be risk neutral in \cite{DeatonLaroque} and the key decision is whether product (either from last year or newly harvested) is to be (a) placed on the market for consumption or (b) stored instead. Product will be stored whenever expected future price is sufficiently high relative to the current price. Otherwise product will be sold for current consumption. Market forces, i.e., arbitrage activities, will generally ensure that the allocation between storage and current consumption is such that prices align and commodity owners are indifferent between the sell or store choice.


However, a corner solution may arise in the event of a sequence of poor harvests or alternatively when one very poor harvest occurs and little stored commodity is available. Then the demand for the product right now may be such that no commodity is placed in storage, a situation that is referred to as stock-out. \cite{DeatonLaroque} use the i.i.d. random harvest assumption together with operator theory to establish that the relationship between available stocks (from storage or current harvest) and price is as given by the solid, continuous, two-part curve in Figure \ref{Figure2}. The vertical red arrow shows how storage acts to buttress price when supplies are large. Rather than take a low price on the saturated market for current consumption, stocks are stored in the expectation that prices will rise in a year or two. When a stock-out occurs then price is determined entirely by the current demand curve. However, when stocks exceed a critical level and price is lower than a corresponding critical price then additional stocks available due to a larger harvest are split into two uses. These are additional current consumption and additional storage. 

One interpretation of Figure \ref{Figure2} is to view the horizontal axis as consumption but then recognize that what you see is not what you get because when prices are low then not all additional harvest translates into additional stocks and the price decline is ameliorated. In shifting up the price to stocks curve when stocks are high, the presence of storage should reduce the adverse effect of higher harvest yield on price. Intuitively, this implies that
\begin{subequations}
\begin{align}
    \text{Cor}(y_{jt},p_t | \tilde{s}_t,l) & \leq 0 \text{, and} \label{condcor} \\
    \frac{d\text{Cor}(y_{jt},p_t | \tilde{s}_t,l)}{d\tilde{s}_t} & \geq 0, \label{dcondcor}
\end{align}
\end{subequations}
where $y_{jt}$ is the harvested crop yield for county $j$ in year $t$, $p_t$ is the year $t$ harvest price,  $\tilde{s}_t = \frac{s_t}{\tilde{x}_{t-1}}$ is the amount of production-normalized carryover stocks from year $t-1$ to year $t$\footnote{We normalize by production because when typical production and consumption is larger then more stocks are needed to protect against the costs of future production shortfalls. Corn and soybean production have both expanded over time as the human population has increased, demand for meat has strengthened, and more corn has been diverted to produce biofuels.}, $s_t$ is the actual amount of carryover stocks in from year $t-1$ to year $t$, and $\tilde{x}_{t-1}$ is a localized estimated scatterplot smoothing (LOESS) regression estimate of year $t-1$'s national yield. 
We also view the correlation as being influenced by distance from the core production region as measured by $l$. Yields in any year are positively spatially correlated \citep{Goodwin2001,GongHennessyFeng} because soils and weather events are positively spatially correlated. Therefore any core or central production region will have a large impact in determining price so that price-yield correlations should be most strongly negative in core production regions. This natural hedge in turn generates a comparative advantage in risk management for these regions because low prices and low yield are unlikely to occur simultaneously. In our empirical data study, we will confirm this natural hedge by showing that states in the core production region have large negative price-yield correlations for low stocks and less negative correlation for higher stocks.  \change{States outside the core production region have price-yield correlations that are smaller in absolute value and largely constant across stock levels, while more northernly states have correlations close to zero.}

Also of relevance to us from \cite{DeatonLaroque} is
\begin{align}
    \frac{d\text{Var}(p_t | \tilde{s}_t)}{d\tilde{s}_t} & \leq 0. \label{dcondvar}
\end{align}
Variability of price conditional on stocks will decline with an increase in stocks because when stocks are low then additional current-year yield is consumed and so is devoted entirely to reducing current scarcity. When stocks are high then some of the addition is stored and not marketed, lessening the impact on current price. This point is relevant for commodity markets because one price variability metric is implied volatility, as often extracted from applying the Black formula for options market prices on commodity futures \citep{Black1976}. It is well-known that stocks carried over from prior years decrease futures price volatility \citep{HennessyWahl,KaraliPower}. Our empirical data analysis will provide further evidence that as stocks increase the volatility of the futures price decreases. 


\subsection{Setting Premium Rates and Implication for Indemnities}

Private sector crop insurance offerings have been commercially available for decades in the United States \citep{Gardner}, but have generally not been popular due to high administration costs and more recently due to competition from federal programs. Motivated by concerns about market failures \citep{Chambers,Justetal}, political considerations \citep{Innes} and behavioral attitudes that depress demand \citep{Fengetal,Caietal2020}, \Copy{R1C10}{\change{government support for agricultural insurance is generally large}}. These supports take the forms of paying for infrastructure required to assess risks and of providing subsidies on premiums charged by insurance agents. Experiences in many countries and for many crops have shown that both intervention types are central to developing the levels of sustained grower participation in crop insurance that policy-makers seek \citep{Krameretal}. The product development path traversed by first the United States and more recently, India, China, European Union countries and elsewhere is not unexpected. Take-up is initially limited so contract design and pricing issues are repeatedly revisited, sometimes drawing in more growers and sometimes not \citep{MingYeShi,SanteramoFord,SmithGlauber,Cariappa}. 

While impressive strides have been made in improving rate-setting in the United States \citep{Cobleetal}, problems remain \citep{Ramirezetal,PriceYuHennessy}. A common complaint, which we refer to as spatial adverse selection, is that rate structures are such that farmers operating better quality land within a county are asked to cross-subsidize those owning worse quality land under the same crop in that county \citep{Ramirezetal2017,PriceYuHennessy}. \red{Other sources of variabilities, such as year-to-year weather effects, can also be problematic \citep{WeatherSplinesPaper} as they may be visible and yet not factored into rate-setting processes.} The subject of our investigation is accumulated stocks carried from year $t-1$ to year $t$, where most of the main crops in plant agriculture can be stored across the typical harvest period at an acceptable cost. 

The presence of stocks affects the joint distribution of price and yield, thus impacting the setting of insurance premiums. 
In functional form, the indemnity $I_{jt}$ for a county $j$ in year $t$ is
    $I_{jt} = \max\left(\psi\bar{p}_{t}\bar{y}_t - p_ty_{jt},0\right)$, 
a decreasing and convex function of revenue ($p_ty_{jt}$) where $\bar{p}_{t}$ is the year $t$ spring futures price, $\bar{y}_t$ is the actual production history (APH) yield average for year $t$, the arithmetic average of the past ten years of yield data, as required by federal regulation when available \citep{Cobleetal}, and $\psi$ is the coverage rate of the insurance policy. The premium or expected indemnity payout for county $j$ in year $t$ can be calculated as
\begin{align}
    Premium_{jt} = E[I_{jt}] = \int\max\left(\psi\bar{p}_{t}\bar{y}_t - p_ty_{jt},0\right) dG(y_{jt},p_t)\label{uncondpremium}
\end{align}
where $G(y_{jt},p_t)$ is the joint cumulative distribution function for $y_{jt}$ and $p_t$ with associated density function $g(y_{jt},p_t)$.

\change{In order to clarify matters and lay out our approach to simulations, we have devised several flow graphs presented in Figure \ref{PremMethods}. Panel (a) describes the current RMA approach to modeling revenue distributions. Here pre-season futures price and implied volatility, or volatility factor, determine the harvest price distribution. Stocks are not considered directly in this model, but they do enter in the background. We refer to how they do so as the two-channel model. In it stocks affect futures contract prices because (Channel 1) market participants understand that tight stocks early in the year likely lead to restricted supply at harvest time. Stocks also affect forward-looking volatility (Channel 2) because when stocks are exhausted then they cannot be drawn down in the event that either a poor harvest or, for some reason, a dramatic increase in demand occurs. In our empirical inquiry, we argue that the harvest price distribution depends on stocks, $\tilde{s}_t$, directly through futures prices and implied volatilities. As such, we estimate the futures price and implied volatility by the observed levels of stocks via our two-channel methodology in panel (b) of Figure \ref{PremMethods}.}

\Copy{R1C12}{\change{Others have raised concerns about the bias in correlation statistics entering rate-setting, \citep[see][]{Andersonetal2013}. Our concern is more specific. Separate from calculation methodologies, the correlation will vary with commodity abundance. In our empirical inquiry, we argue that $\text{Cor}(y_{jt},p_t) = \frac{\text{Cov}(y_{jt},p_t)}{\sqrt{\text{Var}(y_{jt})\text{Var}(p_t)}}$ depends on  $\tilde{s}_t$ and location, and should be calculated with respect to a location-specific conditional density, $g(y_{jt},p_t|\tilde{s}_t,c_j)$, where $c_j$ is the state in which county $j$ is located. 
In particular, we seek to characterize how the conditional correlation changes with stocks and location (i.e., state).}} 
\change{Our proposal is illustrated in panel (c) of Figure \ref{PremMethods}, flowing from stocks directly to yield. As we will explain, our rationale is that yields are temporarily autocorrelated due to temporal autocorrelation in one or both of weather patterns or carryover moisture available. This channel would have observable stocks as a signal for both the i) yield distribution and ii) price-yield correlation.} 
If these effects are not fully accounted for in the insurance premium determination formula, then potential heterogeneity issues may impede insurance program participation and/or require additional subsidies to overcome. In addition, as the correlation is likely to be most negative in years with high price and low production, the stabilization effect of the natural hedge may not be adequately acknowledged by any insurance pricing formula. \Copy{R1C11}{\change{This inference \red{aligns with the perspective} 
that core production regions may face premiums that are too high in comparison to long run indemnities incurred, a perspective that is supported by empirical study of insurance program performance \citep{USGAO2015,ChenSherrick}.}} However, no relevant financial instrument is available for correlation and no effort is made to condition correlation on market information. \red{Therefore, the potential for bias inherent in traditional premium calculation methods arises whenever equations (\ref{condcor}) and (\ref{dcondcor}) are applicable.} 

\section{Methodology} \label{Methodology}

Define $g(p_t | \tilde{s}_t)$ and $g(y_{jt} | p_t, \tilde{s}_t,c_j)$ as the density functions of $p_t$ conditioning on $\tilde{s}_t$ and of $y_{jt}$ conditioning on $p_t$, $\tilde{s}_t$, and $c_j$, respectively, where it follows that $g(y_{jt},p_t | \tilde{s}_t,c_j) = g(p_t | \tilde{s}_t)g(y_{jt} | p_t, \tilde{s}_t,c_j)$. Because $p_t$ is a national harvest price, $g(p_t | \tilde{s}_t)$ does not depend on state $c_j$. For the sake of simplicity, we exclude $c_j$ from the condition and fit the model separately for different states. For $\tau_{p} \in (0,1)$ and $\tau_y \in (0,1)$, let $q_{\tau_p}(\tilde{s})$ and $q_{\tau_y}(p,\tilde{s})$ denote the $\tau_p^{th}$- and $\tau_y^{th}$- conditional quantile functions of $g(p| \tilde{s})$ and $g(y|p,\tilde{s})$, respectively. 
Using semi-parametric quantile regression, we can estimate these possibly non-linear quantile functions as $\hat{q}_{\tau_p}(\tilde{s})$ and $\hat{q}_{\tau_y}(p,\tilde{s})$, respectively. Therefore, if $\tau_{p,r}$ and $\tau_{y,r}$ $(r = 1,2,\cdots,R)$ are independently drawn from the uniform distribution on $(0,1)$, then $p_r^* = \hat{q}_{\tau_{p,r}}(\tilde{s})$ and $y_r^* = \hat{q}_{\tau_{y,r}}(p_r^*,\tilde{s})$ provide independent random samples from the conditional joint density $g(y,p|\tilde{s})$. 

Multiple choices exist for estimating ${q}_{\tau_p}(\tilde{s})$ and ${q}_{\tau_y}(p,\tilde{s})$, such as kernel weighting methods \citep{YuandJones} and smoothing spline methods \citep{Portnoy1994}. Among all the literature findings, one important conclusion is that there exists a significant tradeoff between computational efficiency and smoothness of the regression estimates. Specifically, while unsmoothed quantile regression methods are computationally efficient, they can produce spiky distributional curves. On the other hand whereas smoothed regression methods produce a smoother distributional curve, they come at increased computational cost for calculating the ``best'' smoothing value. For this paper, we employ a quantile regression method based on penalized B-splines \citep{Yoshida2013,ChenandYu}. This approach provides a relatively smoothed quantile function that comes at reduced computational burden, without specifying a form for the quantile function. 



\subsection{SQR with Penalized B-Splines} \label{QuantReg}
For a general $\boldsymbol{x} = \{x_{1},x_{2},\cdots,x_{d}\}^T$, $d\times1$ vector of observed covariates, define the $\tau^{th}-$quantile function as $q_{\tau}(\boldsymbol{x}) = \boldsymbol{B}^T(\boldsymbol{x})\boldsymbol{\beta}_{\tau}$ where $\boldsymbol{B}(\boldsymbol{x}) = \{\boldsymbol{B}_1^T(x_{1}),\cdots,\boldsymbol{B}_d^T(x_{d})\}^T$ 
\red{and $\boldsymbol{B}_l(x_l)$ for $l = 1,\cdots,d$ is the B-spline basis function that converts a scalar $x \in \chi_l$ to
a $(r+K_n) \times 1$ vector in $\mathbb{R}^{r+K_n}$;}
 $r$ and $K_n - 1$ are the degree and number of knots of the B-spline basis, respectively; and $\boldsymbol{\beta}_{\tau}$ is the \change{$d(r+K_n) \times 1$} vector of the $\tau^{th}-$quantile regression coefficients. 
 \red{B-spline offers a relatively smooth quantile regression function without requiring a specific
 parametric assumption for the conditional distribution. 
More specifically, 
the quantile function at specific $\tau$ values can take on diverse forms, enabling our SQR method to effectively capture any non-linear mean structure, skewness and heteroskedasticity in the conditional distribution. }
This flexibility empowers us to refrain from imposing rigid assumptions regarding the shape of the distribution.

Details about the construction of the B-spline bases can be found in Section \ref{BsplineCalc} of the online supplement, and readers can learn more properties that splines possess in \cite{Splines}. The estimated quantile regression function is then defined as 
\begin{align}
    \hat{q}_\tau(\boldsymbol{x}) & = \boldsymbol{B}^T(x)\hat{\boldsymbol{\beta}}_\tau \nonumber, \text{ where} \\
    \hat{\boldsymbol{\beta}}_{\tau} & = \argmin_{\boldsymbol{\beta}} \sum_{i=1}^n \rho_\tau\left(y_i - \boldsymbol{B}^T(\boldsymbol{x}_i){\boldsymbol{\beta}}\right) + \frac{\lambda}{2}{\boldsymbol{\beta}}^T\boldsymbol{D}_m^T\boldsymbol{D}_m{\boldsymbol{\beta}} \label{minimizer}.
\end{align}
Here $\lambda (> 0)$ is the quantile regression smoothing parameter, and $\boldsymbol{D}_m$ is the $m^{th}$ difference matrix \citep[see][among others]{PriceYuHennessy}. 
When $m=2$, $\boldsymbol{D}_m$ has an interpretation related to the integral of the square of the second derivative of the function defined by the B-spline.
$\boldsymbol{D}_2$ is the choice of penalty matrix we use for the rest of this discussion, \change{while the $\lambda$, $K_n$, and $r$ choices will be discussed in Section \ref{Empirical}}.
The method is implemented to estimate two quantile functions $\hat{q}_{\tau_p}(\tilde{s})$ and $\hat{q}_{\tau_y},(p,\tilde{s})$ 
 which are used to generate draws from $g(y,p|\tilde{s})$.

\subsection{Sampling from the conditional distribution $g(y,p|\tilde{s})$}\label{Sampling}

To correctly examine the impact of stocks on correlation and insurance premiums by using aggregated data from multiple years, we must address two issues. Firstly, there may exist time-dependent trends between the county-level yields and national harvest price in the current year and the observations from previous years. Secondly, the harvest price has not been adjusted for inflation, which can potentially beget misleading conclusions regarding the influence of stocks. 

To address the first issue, we mitigate by detrending each of these variables. This involves converting $y_{jt}$ and $p_t$ to detrended variables, denoted as 
\red{$\tilde{y}_{jt}=y_{jt}-\hat{y}_{c_j,t}$ and $\tilde{p}_t=\log p_t-\hat{p}_t$, }
respectively, where $\hat{y}_{c_j,t}$ and $\hat{p}_t$ are the estimated trends. 
We then apply our method on these detrended variables to obtain samples
$\{\tilde{y}_{r}^*,\tilde{p}_{r}^*\}_{r=1}^R$ 
from the distribution of detrended yield and price, represented as $g(\tilde{y},\tilde{p}|\tilde{s})$.
\change{In addition, we discuss a method to obtain samples 
\red{$\{\hat{y}_{r,t}^*,\hat{p}_{r,t}^*\}_{r=1}^R$}  
from the sampling distribution of the trends, $\hat{y}_{c_j,t}$ and $\hat{p}_t$.} 
\change{We use these sampled trends to perform retrending and obtain samples 
\red{$\{y_{r,t}^*,p_{r,t}^*\}_{r=1}^R$} from the distribution $g({y},{p}|\tilde{s}_t)$.} 
To tackle the second issue, we adjust the samples \red{$\{y_{r,t}^*,p_{r,t}^*\}_{r=1}^R$} using the GDP deflator for prices and the estimated yield trends for yields. This adjustment allows us to obtain the adjusted samples $\{y_{a,r,t}^*,p_{a,r,t}^*\}_{r=1}^R$, where the yields and prices are expressed in units corresponding to year $a$. Section \ref{InflationAdj} in the online supplement discusses the method used to convert the price $p_t$ and county-level yield $y_{jt}$ to units corresponding to year $a$. In this paper, we have chosen to set $a = 2020$. Now we will provide a description of our sampling procedure.

\Copy{R1C15}{\change{For $p_t$, we perform a penalized B-spline regression on
$\log(p_t)$ to estimate the trend $\hat{p}_t$.  For simplicity, the estimated trend as well as the detrending price $\tilde{p}_t$ are both presented on the log-scale. Define a B-spline basis for this regression with $r=3$, $K_n = \lceil \frac{T}{10} \rceil$, $\chi_t = [0,T]$, knots evenly spaced every 10 years, and $\lambda$ is chosen via a generalized approximate cross validation (GACV) method \citep{GACV}.\footnote{For $y_i$ and $\boldsymbol{x}_i = \{x_{1i},x_{2i},\dots,x_{di}\}^R$,  $i = 1,\dots,n$, $\hat{\lambda} = \argmin_{\lambda} \frac{\sum_{i=1}^n \left(y_i - \boldsymbol{B}^T(\boldsymbol{x}_i){\boldsymbol{\beta}}\right)^2}{n - tr(\boldsymbol{H})}$ where $\boldsymbol{H}$ is the so-called ``hat matrix'' for the penalized loss function $\sum_{i=1}^n \left(y_i - \boldsymbol{B}^T(\boldsymbol{x}_i){\boldsymbol{\beta}}\right)^2 + \frac{\lambda}{2}{\boldsymbol{\beta}}^T\boldsymbol{D}_m^T\boldsymbol{D}_m{\boldsymbol{\beta}}$.} We estimate the year $t$ trend $\hat{p}_{t}$ to be
\begin{align}
    \hat{p}_{t} & = \boldsymbol{B}^T(t)\hat{\boldsymbol{\beta}}_p, \nonumber \\
    \hat{\boldsymbol{\beta}}_p & = \argmin_{\boldsymbol{\beta}_p} \sum_{t=1}^T \left(\log(p_t) - \boldsymbol{B}^T(t)\boldsymbol{\beta}_p\right)^2 + \frac{\lambda}{2}{\boldsymbol{\beta}_p}^T\boldsymbol{D}_2^T\boldsymbol{D}_2{\boldsymbol{\beta}_p}, \label{detrend-p}
\end{align}
and the detrended price, $\tilde{p}_{t}$, to be
\begin{align}
    \tilde{p}_t = \log(p_t) - \hat{p}_t. \label{detrendedp}
\end{align}}

\change{For $y_{jt}$, we perform a penalized B-spline regression for mean yearly crop yield by state using the same B-spline basis and smoothing parameter selection method as in (\ref{detrend-p}). We estimate the year $t$ trend $\hat{y}_{t}$ to be
\begin{align}
    \hat{y}_{t} & = \boldsymbol{B}^T(t)\hat{\boldsymbol{\beta}}_y, \nonumber \\
    \hat{\boldsymbol{\beta}}_y & = \argmin_{\boldsymbol{\beta}_y} \sum_{t=1}^T \sum_{j=1}^{n_t} (y_{jt} - \boldsymbol{B}^T(t)\boldsymbol{\beta}_y)^2 + \frac{\lambda}{2}{\boldsymbol{\beta}_y}^T\boldsymbol{D}_2^T\boldsymbol{D}_2{\boldsymbol{\beta}_y}, \label{detrend-y}
\end{align}
where $n_t$ is the number of county-level data points in the given state. The detrended county-level yield, $\tilde{y}_{jt}$, is
\begin{align}
    \tilde{y}_{jt} = y_{jt} - \hat{y}_{t}. \label{detrendedy}
\end{align}}

\change{Using the asymptotic properties from \cite{Claeskensetal2009}, 
\red{
we simulate the random trends as follows:
\begin{align}
    \hat{p}_{r,t}^* \overset{.}{\sim} \mathcal{N}\left(\hat{p}_t,\frac{\hat{\sigma}_p^2}{T}\boldsymbol{B}^T(t)\hat{G}^{-1}\boldsymbol{B}(t)\right), \nonumber \\
    \hat{y}_{r,t}^* \overset{.}{\sim} \mathcal{N}\left(\hat{y}_t,\frac{\hat{\sigma}_y^2}{n}\boldsymbol{B}^T(t)\hat{G}^{-1}\boldsymbol{B}(t)\right) \label{approx_trend_dist}, 
\end{align}
where $n = \sum_{t=1}^T n_t$, 
\begin{align*}
    \hat{\sigma}_p^2 & = \frac{1}{T-1} \sum_{t=1}^T \tilde{p}_t^2, \\
    \hat{\sigma}_y^2 & = \frac{1}{n-1} \sum_{t=1}^T \sum_{j=1}^{n_t} \tilde{y}_{jt}^2, \text{ and} \\
    \hat{G} & = \frac{1}{T} \sum_{t=1}^T \boldsymbol{B}(t)\boldsymbol{B}^T(t).
\end{align*}
}
}

}

Using SQR with penalized B-splines, we obtain the estimated quantile functions for detrended prices and yields as 
\begin{align}
    \hat{q}_{\tau_p}(\tilde{s}) & = \boldsymbol{B}^T(\tilde{s}) \hat{\boldsymbol{\beta}}_{\tau_p} \label{quantreg_ptilde} \text { and} \\
    \hat{q}_{\tau_y}(\tilde{p},\tilde{s}) & = \boldsymbol{B}^T(\tilde{p}_t,\tilde{s}_t) \hat{\boldsymbol{\beta}}_{\tau_y},\label{quantreg_ytilde}
\end{align}
where $\boldsymbol{B}(\tilde{p},\tilde{s}) = \{\boldsymbol{B}^T(\tilde{p}),\boldsymbol{B}^T(\tilde{s})\}^T$. Here $\boldsymbol{B}(\tilde{p}_t)$ and $\boldsymbol{B}(\tilde{s}_t)$ are constructed using a method outlined in Section \ref{BsplineCalc} of the online supplement;
\begin{align}
    \hat{\boldsymbol{\beta}}_{\tau_p} &= \argmin_{\boldsymbol{\beta}} \sum_{t=1}^T \rho_{\tau_p}\left(\tilde{p}_t - \boldsymbol{B}^T(\tilde{s}_t){\boldsymbol{\beta}}\right) + \frac{\lambda}{2}{\boldsymbol{\beta}}^T\boldsymbol{D}_2^T\boldsymbol{D}_2{\boldsymbol{\beta}}\label{betahat_p}\text{; and} \\
    \hat{\boldsymbol{\beta}}_{\tau_y} &= \argmin_{\boldsymbol{\beta}} \sum_{t=1}^T\sum_{j=1}^{n_{t}} \rho_{\tau_y}\left(\tilde{y}_{jt} - \boldsymbol{B}^T(\tilde{p}_t,\tilde{s}_t){\boldsymbol{\beta}}\right) + \frac{\lambda}{2}{\boldsymbol{\beta}}^T\boldsymbol{D}_2^T\boldsymbol{D}_2{\boldsymbol{\beta}}\label{betahat_y}.
\end{align}
\Copy{R2C5}{\change{In our empirical data study, for the B-spline bases we select tuning parameters $r = 3$ and $K_n = 4$, while the smoothing parameter, $\lambda$, is chosen via GACV.\Copy{R1C19}{\footnote{For $y_i$ and $\boldsymbol{x}_i = \{x_{1i},x_{2i},\dots,x_{di}\}^R$,  $i = 1,\dots,n$, $\hat{\lambda} = \argmin_{\lambda} \frac{\sum_{i=1}^n \rho_\tau\left(y_i - \boldsymbol{B}^T(\boldsymbol{x}_i){\boldsymbol{\beta}}\right)}{n - tr(\boldsymbol{H})}$ where $\boldsymbol{H}$ is the so-called ``hat matrix'' for the penalized loss function $\sum_{i=1}^n \rho_\tau\left(y_i - \boldsymbol{B}^T(\boldsymbol{x}_i){\boldsymbol{\beta}}\right) + \frac{\lambda}{2}{\boldsymbol{\beta}}^T\boldsymbol{D}_m^T\boldsymbol{D}_m{\boldsymbol{\beta}}$.  For ease of use in our empirical study, we minimize $\hat{\lambda}$ across all values of $\tau_p$ and $\tau_y$.}}}}

For a given year ($t$) with $\tilde{s}_t=\tilde{s}$,  $R$ independent samples from the joint distribution of detrended yields and prices are obtained as $\{\tilde{p}_r^* = \hat{q}_{\tau_{p,r}}(\tilde{s}), \tilde{y}_r^* = \hat{q}_{\tau_{y,r}}(\tilde{p}_r^*,\tilde{s})\}_{r=1}^R$, where $\{\tau_{p,r}\}_{r=1}^R$ and $\{\tau_{y,r}\}_{r=1}^R$ are randomly sampled from $Uniform(0,1)$. 
\change{We randomly sample the trends using (\ref{detrend-p}), (\ref{detrend-y}), and (\ref{approx_trend_dist}) 
Based on (\ref{detrendedp}) and (\ref{detrendedy}), we perform retrending as 
\red{$p^*_{r,t} = \exp(\hat{p}^*_{r,t} + \tilde{p}^*_{r,t})$ and $y_{r,t}^* = \hat{y}^*_{r,t} + \tilde{y}_{r,t}^*$. Then $\{p_{r,t}^* , y_{r,t}^* \}_{r=1}^R$} are further converted into the year 2020 units, i.e., 
$\{p_{2020,r,t}^* , y_{2020,r,t}^* \}_{r=1}^R$, according to the adjustment procedure described in Section \ref{InflationAdj} of the online supplement. }We then use these samples to obtain Monte Carlo approximations of the conditional correlation in (\ref{condcor}) and conditional insurance premium in (\ref{uncondpremium}).  \Copy{R1C20}{\change{The efficacy of our proposed methodology is verified in a thorough simulation study in Section \ref{Simulation} of the online supplement.}}

\section{Empirical Study} \label{Empirical}

In this section, we use our method to generate samples from both the stock-conditioned joint density $g(\tilde{y},\tilde{p}|\tilde{s})$ and the unconditional joint density $g(\tilde{y},\tilde{p})$, which is independent of stocks. These samples are then employed to compute the crop insurance premium as well as the correlation between harvest price and county-level yield. The methodology for sampling from $g(\tilde{y},\tilde{p})$ is a special case of our sampling methodology when drawing from $g(\tilde{y},\tilde{p}|\tilde{s})$, where details are provided in Appendix \ref{unconditional} of the on-line supplement. 
\Copy{R3C3}{\change{For this data analysis, we obtain data on corn and soybeans from twelve U.S. states which are considered to be part of the ``Corn Belt'' (Illinois, Indiana, Iowa, Kansas, Michigan, Minnesota, Missouri, Nebraska, North Dakota, Ohio, South Dakota, and Wisconsin) from the years 1990 to 2018 (i.e., $T = 29$) using National Agricultural Statistics Service (NASS) survey data supported by the US Department of Agriculture.  To control for irrigation, we only use non-irrigated yield results for Nebraska and Kansas.  We obtain harvest time and February futures price (in U.S.\$ per bushel) for corn and soybeans, national-level yield and end-of-year carryover stocks for corn and soybeans, and state and county-level yields for corn and soybeans.} Note that each state has a different number of counties and not every county has a reported yield in each year (i.e., $n_t$ is different for $t = 1,\cdots,T$). \change{In total we have $29$ observed values of $p_t$ and $\tilde{s}_t$ for $t = 1,\cdots,T$ for both corn and soybeans, with $26,666$ and $24,075$ observed values of $y_{jt}$ for corn and soybeans, respectively, $t = 1,\cdots,T$, and $j = 1,\cdots,n_t$.} Descriptive statistics for corn and soybeans can be found in Table \ref{tbl:Corn_desc}.}

\spacingset{2}
For each state we apply the detrending procedures outlined in (\ref{detrendedp}) and (\ref{detrendedy}) to obtain $\tilde{p}_t$ and $\tilde{y}_{jt}$ for $t = 1,\cdots,T$ and $j = 1,\cdots,n_t$. These detrended values are then used in our SQR methodology to generate samples  $\{\tilde{y}_{r}^*,\tilde{p}_{r}^*\}_{r=1}^R$ from both $g(\tilde{y}_{jt},\tilde{p}_t|\tilde{s}_t)$ and $g(\tilde{y}_{jt},\tilde{p}_t)$, respectively. \change{We also randomly sample the trend using (\ref{detrend-p}), (\ref{detrend-y}), and (\ref{approx_trend_dist}) to obtain $\{\hat{y}_{r}^*,\hat{p}_{r}^*\}_{r=1}^R$.  We then apply our retrending procedure outlined in the previous section as well as the inflation adjustment from Section \ref{InflationAdj} of the online supplement to obtain samples from $\{{y}_{2020,r}^*,{p}_{2020,r}^*\}_{r=1}^R$ from both $g({y}_{2020,jt},{p}_{2020,t}|\tilde{s}_t)$ and $g({y}_{2020,jt},{p}_{2020,t})$.}
For the empirical data study we set $R = 1000$. The corn and soybean results are empirically similar, so we only present the corn results in this paper; the soybean results can be found in Appendix \ref{SoybeanResults} of the on-line supplement. 

\change{Figure \ref{Corn_Var} presents the estimated standard deviation of $p_{2020,t}$ across different levels of stocks from the conditional density function.  We also include a smoothed estimate of the standard deviation from a LOESS regression represented by a solid black line.
As discussed in the introduction, we want to understand the relationship between stocks and price variability to help understand the source of large spikes in commodity prices. We observe that the conditional standard deviation tends to decrease as level of stocks increases, as theorized in (\ref{dcondvar}), suggesting that large and dramatic price spikes are more likely for low levels of stock, and tends to become less likely as stocks increase.}


Figure \ref{Cor_corn_state} plots visualizations of the impact of stocks on the correlation between the harvest time price and county-level yields (in 2020 units) for corn in each of the twelve states. \change{Recall, we only use non-irrigated yields for Kansas and Nebraska.}
Because of the size of the graph, we only include a smoothed estimate of the conditional (black solid curve) and unconditional (dashed red curve) correlations from LOESS regression. In addition to the estimates of the conditional correlations as functions of stocks, we display a 95\% confidence band for the conditional correlation estimate, shown as black dotted lines. 
\Copy{R2C5a}{\change{The standard error of the estimated conditional standard deviation is computed using a Jackknife delete-a-group method with $B = 50$ groups. 
Within each state we order the data points by county, and systematically assign price-yield pairs to form a Jackknife group. We run our SQR methodology as outlined in \ref{Sampling} for each jackknife group, including a recalculation of the smoothing parameter, $\lambda$, by GACV. The variance of the conditional standard deviation is then estimated as 
\begin{align}
    \hat{\text{Var}}(\hat{\text{cor}}(y_{2020,jt},p_{2020,t} | \tilde{s}_t)) = \frac{B-1}{B} \sum_{b=1}^B (\hat{\text{cor}}_b - \bar{\hat{\text{cor}}})^2 \label{cor_jk}
\end{align}
where $\hat{\text{cor}}_b$ is the estimate of the conditional price-yield correlation with the $b^{th}$ Jackknife group deleted using the same SQR methodology outlined in Section \ref{Sampling} and $\bar{\hat{\text{cor}}} = \frac{1}{B} \sum_{b=1}^B \hat{\text{cor}}_b.$
}}

\Copy{R1C25}{\Copy{R2C7}{Much as we had initially theorized in (\ref{condcor}), both the unconditional and conditional correlations for corn are negative \change{for the majority of states}, suggesting that when we observe a negative shock in yield for a given year then we should expect a subsequent increase in the harvest price. In addition, we observe a rise in the conditional correlation as the level of stock increases for the majority of states, providing evidence that correlation also increases as the level of stocks increases, as theorized in (\ref{dcondcor}). This also suggests that the shift in the demand curve for high levels of storage for corn as illustrated in Figure \ref{Figure2} is realized empirically.  \change{However, we note that the correlation structure is not as strong for soybeans. Many of these correlations are also statistically significantly different from the unconditional correlations based on the presented confidence bands. \cite{EfronStein1981} proved that the jackknife estimates of variances are biased upward; therefore, the presented confidence bands are actually conservative.}}


The ``I'' states -- Illinois, Indiana, and Iowa, widely understood as the core production region of corn, and displayed in the top panel of Figure \ref{Cor_corn_state} -- as well as more southern states like Missouri exhibit similar trends in conditional correlation relative to the unconditional correlation for corn; that is the conditional correlation is significantly lower than the unconditional correlation for low levels of stock and increases as stock increases. However, states to the north of the core production region such as Michigan, Minnesota, North Dakota and Wisconsin display conditional correlations that are comparatively much smaller in magnitude, or even positive. Curves for these states also tend to be flatter on the $\tilde{s}_t > 0.1$ domain when compared to the “I” states. Because correlation is a nonlinear function of second moments the unconditional correlation can be uniformly below the conditional correlation, i.e., the solid black curve is not required to cut the dashed red curve and that turns out to be the case for Minnesota. Together, these observations empirically illustrate the natural hedge proposed in the introduction whereby states in core production areas have regional comparative advantage in that current-year yield and harvest price are highly negatively correlated when carryover stocks are low, keeping revenue in those regions stable. In regions going away from the center of production, we see that the price-yield correlation is less sensitive to changes in the amount of carryover stocks, meaning revenue is less stable in these areas, leading to a competitive disadvantage. 

This regional competitive advantage can also be seen in Figure \ref{state_map_corn}, where we present the conditional correlations as geographic heat maps for the observed 0.2-, 0.5-, and 0.8-quantiles of leftover stocks for corn, where the states displaying darker shades of red represent states with more negative price-yield correlations. \change{In addition, we plot the current correlation used by the RMA as discussed in \cite{Goodwinetal2014} in the bottom right panel. As illustrated in that plot, the RMA's estimate of the correlation can be missrepresented for year's when stocks are high, especially for states that are in the periphery.}}

\Copy{R1C26}{\change{One possible rationale for the potential differences in the conditional and unconditional density functions for the county-level yield (and subsequent price-yield correlations) is that weather patterns are not independent between consecutive years \citep{DustBowl,WeatherSplinesPaper}.} 
\change{Furthermore, even if they were temporally independent physical conditions may cause autocorrelations in what matters for crop production.  Good soils can store moisture across years so that water availability is positively autocorrelated even though precipitation is not. We test to ascertain whether there is autocorrelation between county-level yields by fitting the observed, detrended county-level yields for both corn and soybeans from (\ref{detrendedy}) and fit them to an AR-1 model for each state, 
\begin{align}
    \tilde{y}_{jt} = \rho_{0,c_j} + \rho_{1,c_j}\tilde{y}_{j,t-1} + \epsilon_{jt}, \label{autocor_model}
\end{align}
where $\epsilon_{jt}$ are normally distributed error terms and $c_j$ is the state in which county $j$ is located. Because we assume a stationary distribution for $\tilde{y}_{jt}$, $\rho_{1,c_j}$ represents the lag-1 autocorrelation between the detrended county-level yields. The estimates of the lag-1 autocorrelation along with their associated 95\% confidence intervals for both corn and soybeans are presented in Table \ref{AutoCorTable}. All of the lag-1 autocorrelations are positive and significantly different than zero, providing further evidence of a weather effect for the distribution of the county-level yields that is incorporated in our methodology by conditioning on the level of leftover stocks.\footnote{\cite{RobertsSchlenker} used past yields to identify commodity supply and demand elasticities. An assumption made by them is that yields are uncorrelated over time. In their study of yields around the world, they find support for this assumption. We argue that Table 1 findings are to be expected because of weather cycles. Others have acknowledged autocorrelation structures in yields \citep[see]{SpatioTemporal,Schubertetal2004}}.}}

\Copy{R1C33}{We further illustrate the natural hedge through estimation of the conditional insurance premium, hereafter referred to as the proposed insurance premium, as defined in (\ref{uncondpremium}).  \change{More specifically, we compare the conditional insurance premiums from our proposed three-channel method to the two-channel method as illustrated in Figure \ref{PremMethods}. To incorporate stocks directly into our premium calculation, we need to fit a model of spring futures price and implied volatility independently by stocks.  We perform a penalized B-spline regression analysis similar to (\ref{detrend-p}) - (\ref{approx_trend_dist}), except now we calculate the B-spline bases using stocks. Using the B-spline bases outlined in Section \ref{Methodology}, we assert that 
\begin{align}
    \hat{\bar{p}}_t & \overset{.}{\sim} \mathcal{N}\left(\boldsymbol{B}^T(\tilde{s}_t){\boldsymbol{\hat{\beta}}}_{\bar{p}},\frac{\hat{\sigma}_{\bar{p}}^2}{T}\boldsymbol{B}^T(\tilde{s}_t)\hat{F}^{-1}\boldsymbol{B}(\tilde{s}_t)\right), \nonumber \\
    \hat{IV}_t & \overset{.}{\sim} \mathcal{N}\left(\boldsymbol{B}^T(\tilde{s}_t){\boldsymbol{\hat{\beta}}}_{IV},\frac{\hat{\sigma}_{IV}^2}{T}\boldsymbol{B}^T(\tilde{s}_t)\hat{F}^{-1}\boldsymbol{B}(\tilde{s}_t)\right) \label{approx_pbar_IV_dist}.
\end{align}
where
\begin{align*}
     \hat{\boldsymbol{\beta}}_{\bar{p}} & = \argmin_{\boldsymbol{\beta}_{\bar{p}}} \sum_{t=1}^T \left(\log(\bar{p}_t) - \boldsymbol{B}^T(\tilde{s}_t)\boldsymbol{\beta}_{\bar{p}}\right)^2 + \frac{\lambda}{2}{\boldsymbol{\beta}_{\bar{p}}}^T\boldsymbol{D}_2^T\boldsymbol{D}_2{\boldsymbol{\beta}_{\bar{p}}}, \\
          \hat{\boldsymbol{\beta}}_{IV} & = \argmin_{\boldsymbol{\beta}_{IV}} \sum_{t=1}^T \left(\log(IV_t) - \boldsymbol{B}^T(\tilde{s}_t)\boldsymbol{\beta}_{IV}\right)^2 + \frac{\lambda}{2}{\boldsymbol{\beta}_{IV}}^T\boldsymbol{D}_2^T\boldsymbol{D}_2{\boldsymbol{\beta}_{IV}}, \\ 
    \hat{\sigma}_{\bar{p}}^2 & = \frac{1}{T-1} \sum_{t=1}^T (\log(\bar{p}_t) - \boldsymbol{B}^T(\tilde{s}_t)\hat{\boldsymbol{\beta}}_{\bar{p}})^2,\\
    \hat{\sigma}_{IV}^2 & = \frac{1}{T-1} \sum_{t=1}^T (IV_t - \boldsymbol{B}^T(\tilde{s}_t)\hat{\boldsymbol{\beta}}_{IV})^2, \text{ and} \\
    \hat{F} & = \frac{1}{T} \sum_{t=1}^T \boldsymbol{B}(\tilde{s}_t)\boldsymbol{B}^T(\tilde{s}_t).
\end{align*}
The insurance premiums for the three-channel method are then simulated as follows:}} 

\spacingset{1}
\Copy{R1C33i}{\change{
\begin{enumerate}[label = {\arabic*.}]
    \item For $r$ in 1 to $R$:
    \begin{enumerate}[label = {(\alph*)}]
        \item Simulate $\bar{p}_r^+ \sim \mathcal{N}(\hat{\bar{p}}_t, \hat{\sigma}_{\bar{p}}^2)$
        \item Simulate $IV_r^+ \sim \mathcal{N}(\hat{IV}_t, \hat{\sigma}_{IV}^2)$
        \item Simulate $p_{2020,r}^+ \sim \mathcal{LN}(\log(\bar{p}_r^+),IV_r^+)$
        \item Calculate $\tilde{y}_{2020,r}^+ = \hat{q}_{\tau_{y},r}(\log(p_{2020,r}^+) - \hat{p}_r^*,\tilde{s})$
        \item Calculate $y_{2020,r}^+ = \hat{y}_r^* + \tilde{y}_{2020,r}^+$
    \end{enumerate}
    End For
    \item Calculate $Prem_t = \frac{1}{R} \sum_{r=1}^R\max\left(\psi\bar{p}_{r}^+\bar{y}_t - p_r^*y_r^*,0\right)$
\end{enumerate}}}

\spacingset{2}
\Copy{R1C33ii}{\change{Lines (1a)-(1c) outline our methodology to simulate the harvest prices via our two-channel and three-channel models.  Lines (1d)-(1e) is our SQR methodology to simulate the county-level yields as utilized in Section \ref{Methodology} for our three-channel model. To simulate the county-level yields for the two-channel model, we change line (1d) of the above algorithm to read as  $\tilde{y}_{2020,r}^+ = \hat{q}_{\tau_{y},r}(\log(p_{2020,r}^+) - \hat{p}_r^*)$.}

\change{To evaluate our proposed three-channel model against the two-channel model, we perform a rating game as proposed by \cite{KerMcGowan} and used in \cite{Parketal2019}.  The game is first conducted by calculating the loss-ratio for a given year in a given crop-state-coverage combination by
\begin{align}
    LR_{t} = \frac{\sum_{j=1}^{n_t} \max(\phi \bar{p}_t \bar{y}_t - p_t y_{jt},0)}{\sum_{j=1}^{n_t} Prem_t}. \label{LossRatio}
\end{align}
From the perspective of a private insurance marketing company, their proposed premium rate can be compared to that of the rate proposed by the Federal Crop Insurance Corporation (FCIC).  Under the Standard Reinsurance Agreement, the private insurance company is allowed to cede policies to the government. The private company will do so whenever their proposed rate is higher, since the government rates are underestimated and a loss is expected, equivalent to a loss ratio greater than 1.  On the other hand, if the proposed  rate is lower than the government's rate then the policy will be retained by the insurance company because a profit is expected, equivalent to a loss ratio less than 1. \cite{KerTolhurstLiu} noted that, under the Standard Reinsurance Agreement, private insurance companies have a distinct adverse selection advantage over the FCIC.  Because of this, we calculate an index similar to \cite{Parketal2019} as 
\begin{align}
    D = \frac{LR_C^3/LR_R^3}{LR_C^2/LR_R^2}, \label{D}
\end{align}
where $LR_C^3$ and $LR_R^3$ are the cede and retain loss ratios, respectively for our three-channel model whereas $LR_C^2$ and $LR_R^2$ are the cede and retain loss ratios, respectively for our two-channel model.}

\change{We then perform a hypothesis test similar to \cite{KerTolhurstLiu} where the null hypothesis is that the two methods are identical.  We estimate $D_t$ for $t=1,\dots,29$ for each state for coverage levels $\phi = \{0.7,0.85\}$ and calculate a test statistic, $D^* = \sum_{t=1}^{T} I(D_t > 1)$.  Under our null hypothesis, $D^* \sim \text{Binomial}(T,0.5)$, and our ``p-value'' is calculated as the Type-I error probability estimated from the binomial distribution.  A low p-value provides evidence that the proposed three-channel method is more accurate than the two-channel method, while a high p-value suggests no statistical difference between the two methods.
}

\change{The p-values for the two coverage levels by state for coverage levels $\phi = 0.7$ and $\phi = 0.85$ are presented in Table \ref{tbl::RatingGame_Corn}.  Overall, the three-channel method is not statistically better than the two-channel method, except for Kansas at the 85\% coverage level.  For the ``I'' states and the states in the Corn Belt, the p-values are high, suggesting there is not a statistically significant difference between the two-channel and three-channel methods.} 

\change{We point out that the effect of stocks has already been partially factored into the RMA premium calculation through the futures price and real-time implied volatility that the RMA obtains from options markets. We have accounted for both in our imputation method through the \change{log-normal distribution of harvest prices}. Therefore, we might expect to observe insignificant differences between the imputed current and proposed premiums for levels of stock where the differences between conditional and unconditional correlations are not large. 
We note in passing that our proposed methodology does not take into account any innovations in actuarial rate setting procedures as well as global economic trends, which may also account \change{for significant p-values in Table \ref{tbl::RatingGame_Corn}}. On the whole, while further inquiry is warranted and the constant correlation assumption is not valid, our sense is that the current approach to premium setting is not unsound as a result of this questionable assumption.}} 




\section{Conclusion} \label{Conclusion}
We address whether the inclusion of stocks has a significant impact on correlation between yield and harvest price in the U.S. Corn and soybean markets as well as RI premiums. To answer these questions, we obtain samples from the unconditional joint density function $g(y_{jt},p_t)$ and the conditional joint density function $g(y_{jt},p_t|\tilde{s}_t)$ using SQR with penalized B-splines and then use these samples to obtain approximations for both the correlation and county-level insurance premiums. Our SQR methodology to approximate a joint price-yield density is novel, but accurately estimates the price-yield correlation and revenue insurance premiums. We obtain these samples using county-level yield data and national-level yield, stocks, and price data from both corn and soybean crops across 12 ``corn belt'' states from 1990 to 2018. We observe that lower stocks imply a more significantly negative correlation between yield and price, and also that higher levels of stock typically mean a less negative correlation.

We obtain approximations of the conditional and unconditional correlations broken down by states. We observe that the core cornbelt ``I'' states are more pronounced in how correlation becomes less negative as stocks increase. However, we observe that most of the other states, those in the periphery, do not reveal a significant difference between the conditional and unconditional stocks for the differing levels of stock. \change{Furthermore, the negative correlation is generally weaker in the more northern states.} We also obtain approximations of the state-level insurance premiums using a simulation of the currently imposed methodology as well as our proposed method when conditioning on stocks. \change{We observe most states tend to have an insignificant premium difference using our proposed three-channel methodology compared to the two-channel methodology for all levels of stocks, due in part to the fact that the two-channel method already implicitly incorporates stocks into the premium calculation through two channels.} 

\spacingset{1}

\bibliographystyle{apalike}
\bibliography{jbes-template.bbl}






\begin{table}[htbp!]
    \centering
    \begin{tabular}{|c|cc|cc|}
    & \multicolumn{2}{|c|}{Corn} &  \multicolumn{2}{|c}{Soybean} \\
    \hline
         &  Mean & SD & Mean & SD\\
         \hline
        Leftover Stocks (Million bushels)  &  &  &  & \\
                 \hline
         1990s & 1327 & 512 & 263 & 74.4 \\
         2000s & 1598 & 377 & 252 & 147\\
         2010s & 1509 & 545 & 218 & 108\\
                  \hline
         Harvest Price (\$/acre)&  &  &  & \\
                  \hline
         1990s & 2.46 & 0.435 & 5.99 & 0.775 \\
         2000s & 2.74 & 0.815 & 6.90 & 2.16 \\
         2010s & 4.56 & 1.39 & 11.1 & 2.14 \\
                  \hline
         Harvest Yield (bu/acre)&  &  &  & \\
                  \hline
         1990s & 105 & 31.9 & 35.9 & 8.71 \\
         2000s & 126 & 40.4 & 38.4 & 10.4 \\
         2010s & 146 & 41.6 & 46.3 & 9.99 \\
    \end{tabular}
    \caption{Descriptive statistics for stocks, harvest price and county-level harvest yield by decade for corn crops from 1990-2018.}
    \label{tbl:Corn_desc}
\end{table}


\begin{table}[h!]
\centering
\begin{tabular}{ll}
  \hline
  State & Corn \\ 
  \hline
  Illinois & 0.38 (0.34, 0.41)  \\ 
  Indiana & 0.13 (0.09, 0.17)  \\ 
  Iowa & 0.18 (0.14, 0.22) \\ 
  Kansas & 0.65 (0.62, 0.68)  \\ 
  Michigan & 0.54 (0.50, 0.58) \\ 
  Minnesota & 0.60 (0.57, 0.64)  \\ 
  Missouri & 0.19 (0.15, 0.23) \\ 
  Nebraska & 0.51 (0.48, 0.55) \\ 
  North Dakota & 0.58 (0.53, 0.62)  \\ 
  Ohio & 0.16 (0.12, 0.20) \\ 
  South Dakota & 0.65 (0.61, 0.68) \\ 
  Wisconsin & 0.44 (0.40, 0.48) \\ 
   \hline
\end{tabular}
\caption{Point estimate and 95\% confidence intervals for $\rho_{1,c_j}$ for corn by state.}
\label{AutoCorTable}
\end{table}

\begin{table}[ht]
\centering
\begin{tabular}{lrr}
  \hline
State & 70\% Coverage & 85\% Coverage \\ 
  \hline
Illinois & 0.9693 & 0.8675 \\ 
  Indiana & 0.9693 & 0.7709 \\ 
  Iowa & 0.9999 & 0.9879 \\ 
  Kansas & 0.5000 & 0.0121 \\ 
  Michigan & 0.9693 & 0.9693 \\ 
  Minnesota & 0.2291 & 0.3555 \\ 
  Missouri & 0.9320 & 0.2291 \\ 
  Nebraska & 0.5000 & 0.3555 \\ 
  North Dakota & 0.9693 & 0.5000 \\ 
  Ohio & 0.9879 & 0.8675 \\ 
  South Dakota & 0.3555 & 0.2291 \\ 
  Wisconsin & 0.9320 & 0.8675 \\ 
   \hline
\end{tabular}
\caption{P-values for the premium insurance rating game for testing the three-channel premium rate method against the two-channel method for the non-irrigated corn crop dataset.}
\label{tbl::RatingGame_Corn}
\end{table}

\clearpage

\Copy{R2C3}{\begin{figure}
    \centering
\begin{subfigure}[b]{0.7\textwidth}
         \centering
         \includegraphics[width=\textwidth]{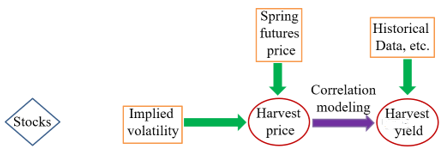}
         \caption{Current RMA Methodology}
     \end{subfigure}
     \hfill
     \begin{subfigure}[b]{0.7\textwidth}
         \centering
         \includegraphics[width=\textwidth]{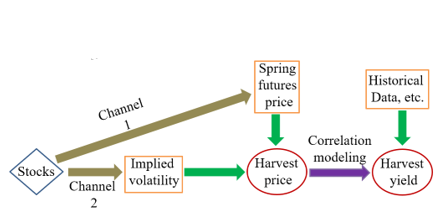}
         \caption{Proposed two-channel Method}
     \end{subfigure}
     \hfill
     \begin{subfigure}[b]{0.7\textwidth}
         \centering
         \includegraphics[width=\textwidth]{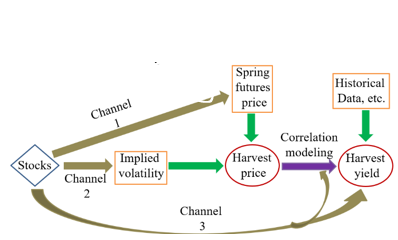}
         \caption{Proposed three-channel Method}
     \end{subfigure}
    \caption{Illustration of the current RMA methodology (top panel), our proposed two-channel method (middle panel), and our proposed three-channel method (bottom panel) for the price-yield distribution used in setting premium rates.}
    \label{PremMethods}
\end{figure}}

\begin{figure}[htbp!]
    \centering
    \includegraphics[scale=0.8]{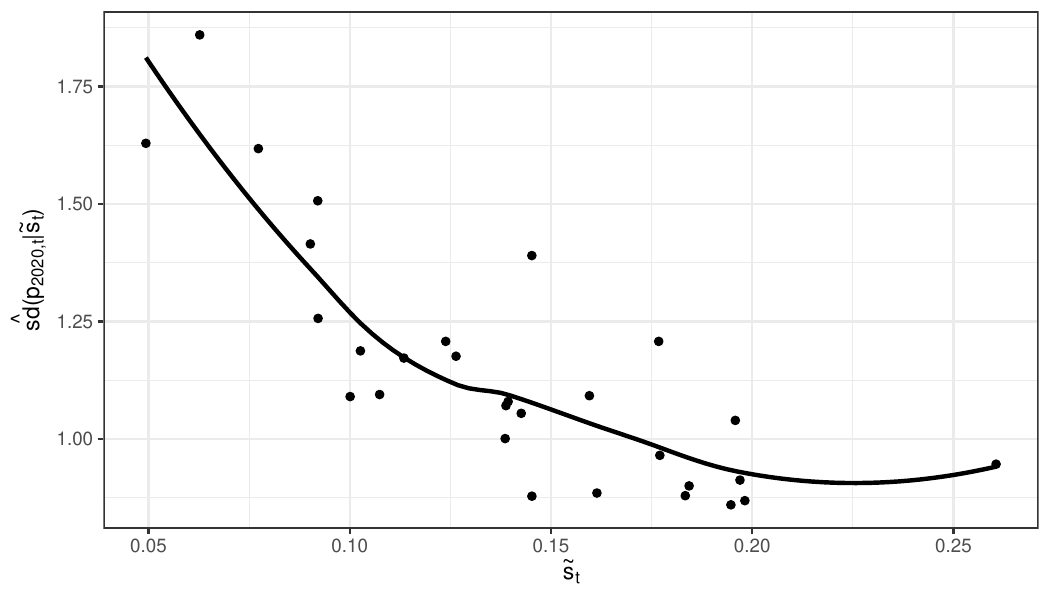}
    \caption{Standard deviation of harvest price as function of stocks for corn crops from 1990-2018 (in 2020 units). The dots represent the estimated standard deviations and the solid black line is the LOESS smoothed regression of the conditional standard deviation function.}
    \label{Corn_Var}
\end{figure}

\begin{figure}[htbp!]
    \centering
    \includegraphics[scale=0.8]{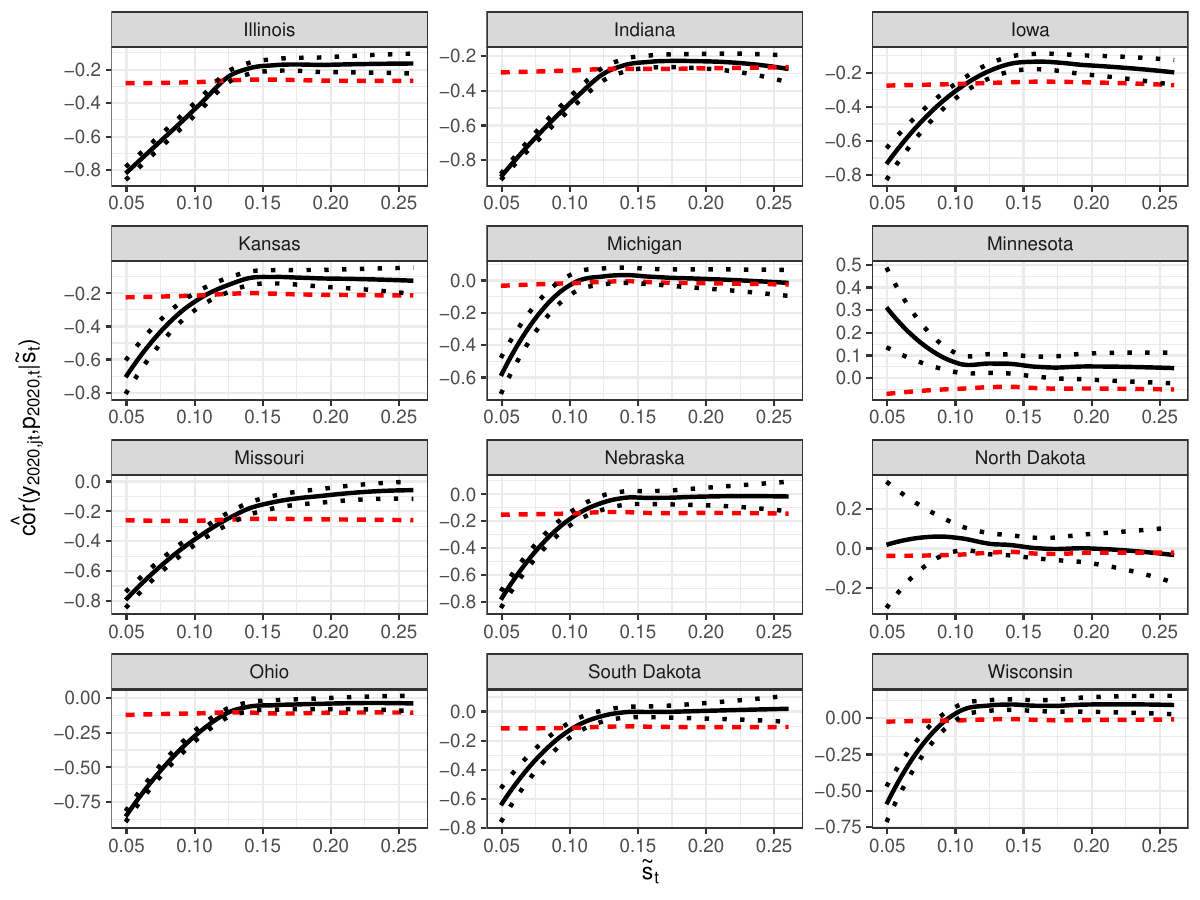}
        \caption{Correlation between harvest price and county-level yield as function of stocks for corn crops by state from 1990-2018 (in 2020 units). The dashed red line is the LOESS smoothed estimate of the unconditional correlation, the solid black line is LOESS smoothed the estimate of the conditional correlation function, and the dotted black lines are the 95\% confidence bands for the conditional correlation function.}
    \label{Cor_corn_state}
\end{figure}

\Copy{R1C34}{
\begin{figure}[htbp!]
    \centering
    \includegraphics[scale=0.9]{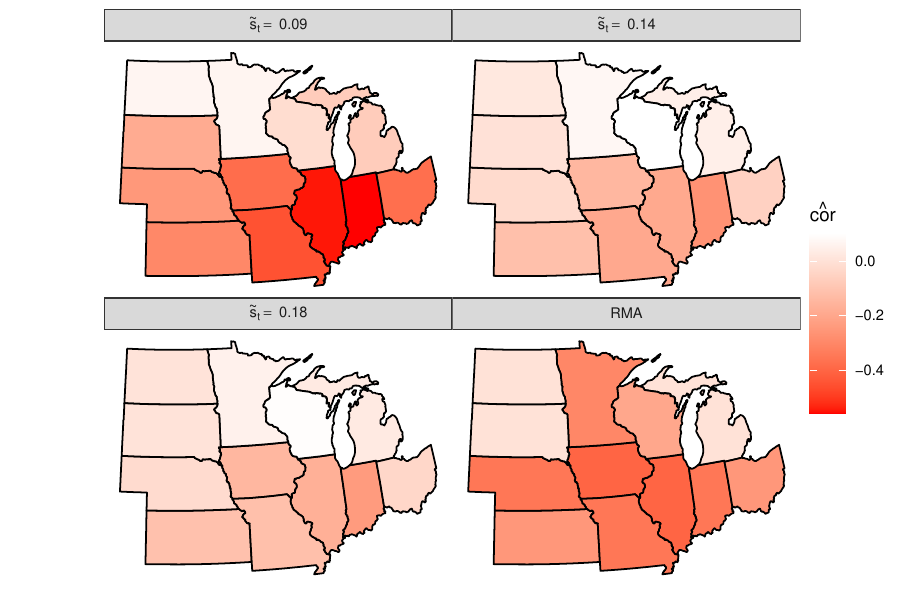}     
    \caption{Heat map of the conditional correlation between harvest price and county-level yield for corn crops across states for the \red{observed 0.2- (top left panel), 0.5-(top right panel), and 0.8- (bottom left panel) quantiles of leftover stocks, corresponding to $\tilde{s}_t=0.09$, $\tilde{s}_t=0.14$ and $\tilde{s}_t=0.18$ respectively. The bottom right corner presents the RMA correlation value from \cite{Goodwinetal2014}.} Darker shades of red represent states with more negative correlation.}
         \label{state_map_corn}
\end{figure}}





\newpage

\numberwithin{equation}{section}
\clearpage
\pagenumbering{arabic}
\renewcommand*{\thepage}{A-\arabic{page}}
\renewcommand*{\thefigure}{A-\arabic{figure}}
\renewcommand*{\thetable}{A-\arabic{table}}
\renewcommand{\thesection}{\Alph{section}}

\setcounter{page}{1}
\setcounter{section}{0}
\setcounter{figure}{0}
\setcounter{table}{0}


\begin{center}
   {\bf Online Supplement for ``The Impact of Stocks on Correlations between Crop Yields and Prices and on Revenue Insurance Premiums using Semiparametric Quantile Regression with Penalized B-splines''}
\end{center}

This online supplement provides additional details about the methodology and simulation results discussed in the main text of ``The Impact of Stocks on Correlations between Crop Yields and Prices and on Revenue Insurance Premiums using Semiparametric Quantile Regression with Penalized B-splines''

\section{B-spline basis formation}
\label{BsplineCalc}

The B-spline basis functions, $\boldsymbol{B}(x)$ for a generic continuous variable, $x$ with degree $p$ and number of knots $K_n - 1$ is constructed as follows:  First, recall that $x \in \chi$ and let us further define $\chi = [l_x,u_x]$ where $l_x$ and $u_x$ are the lower and upper bounds for values of $x$. In order to construct the B-spline basis, we must define knots as $\kappa_k$ for $k = -p,-p+1,\cdots,K_n+p-1,K_n+p$ where $$\kappa_k = \left\{\begin{array}{cc}
   q_{k/K_n}  & 1 \leq k \leq K_n - 1 \\
   l_x & k \leq 0 \\
   u_x & k \geq K_n
\end{array}\right.$$ with $q_{k/K_n}$ being the $\frac{k}{K_n}^{th}-$quantile of the observed values $x_{1},x_{2},\cdots,x_{n}$.

The $p^{th}$ B-spline basis function  is then constructed as
\begin{align}
    \boldsymbol{B}(x) = \left\{ B_{-p+1}^{[p]}(x), B_{-p+2}^{[p]}(x),\cdots,B_{K_n}^{[p]}(x)\right\}' \label{Bspline}
\end{align}
where $B_{k}^{[p]}(x)$ for $k = -p+1,\cdots,K_n$ are defined recursively as follows.

For $s = 0$,
\begin{align*}
    B_{k}^{[0]}(x) = \begin{cases} 1 & \kappa_{k-1} < x \leq \kappa_k \\ 0 & \text{otherwise} \end{cases}\text{, where } k = -p+1,\cdots,K_n +p.
\end{align*}

For $s = 1,2,\cdots,p$,
\begin{align*}
    B_{k}^{[s]}(x) = \frac{x - \kappa_{k-1}}{\kappa_{k+s-1} - \kappa_{k-1}}B_{k}^{[s-1]}(x)+ \frac{\kappa_{k+s}-x}{\kappa_{k+s} - \kappa_{k}}B_{k+1}^{[s-1]}(x), \\
    \text{where } k = -p+1,\cdots,K_n +p - s.
\end{align*}

\section{Inflation Adjustment Procedure}
\label{InflationAdj}

Recall from the main text, we set $\tilde{p}_t$ to be the difference between year $t$'s log-harvest price and a regression estimate of the log harvest price. More specifically, we set
    $\tilde{p}_t = \log(p_t) - \hat{p}_{t}$, 
where $\hat{p}_t$ is a regression estimate of $\log(p_t)$ using localized estimated scatterplot smoothing (LOESS) regressed on year $t$. In addition, for $y_{jt}$, we perform a penalized B-spline regression for mean yearly crop yield by state where $\hat{y}_{c_jt}$ is the year $t$ regression estimate, or trend, for state $c_j$
and the detrended county-level yield, $\tilde{y}_{jt}$, is
    $\tilde{y}_{jt} = y_{jt} - \hat{y}_{c_j,t}$. This section of the online supplement will discuss the methods for going from year $t$'s detrended price and yield ($\tilde{p}_t$, $\tilde{y}_{jt}$), into year $t$'s price and yield expressed in 2020 units ($\tilde{p}_{2020,t}$, $\tilde{y}_{2020,jt}$).

For $p_t$, we account for inflation by dividing $p_t$ by the GDP deflator, which measures the changes in prices for all the goods and services produced in an economy. The GDP deflator for year $t$ is defined as $GDPDEF_t = \frac{Nominal_t}{Real_{2020,t}}$,
where $Nominal_t$ is the U.S. Nominal GDP in year $t$ and $Real_{2020,t}$ is the U.S. Real GDP in year $t$ expressed in 2020 dollars. Thus, we obtain $p_{2020,t}$ as
\begin{align}
    p_{2020,t} = \frac{p_t}{GDPDEF_t}. \label{p2020}
\end{align}
We also need to adjust the county-level yield to be expressed in 2020 units. To go from $\tilde{y}_{jt}$ to $y_{jt}$, we add back $\hat{y}_{c_jt}$, the estimate of annual crop yield in year $t$ in state $c_j$. So to obtain the county-level yield in 2020 units we should add in $\hat{y}_{c_j,2020}$ instead. Thus,
\begin{align}
    y_{2020,jt} & = \hat{y}_{c_j,2020} + \tilde{y}_{jt} \nonumber \\
    & = y_{jt} + \hat{y}_{c_j,2020} - \hat{y}_{c_jt}\label{y2020}.
\end{align}

\section{Simulation Study} \label{Simulation}

In this section, we verify the efficacy of our proposed SQR methodology via a simulation study. It is sufficient  to assess the accuracy of the sampling from $g(\tilde{y},\tilde{p}|\tilde{s})$ for a single state. For $j = 1,\cdots, n_t$ and $t = 1,\cdots,T$, define
    $\tilde{p}_t = \mu_{\tilde{p}}(\tilde{s}_t) + \sigma_{\tilde{p}}(\tilde{s}_t)\epsilon_{t}^p$, and 
    $\tilde{y}_{jt} = \mu_{\tilde{y}}(\tilde{p}_t,\tilde{s}_t) + \sigma_{\tilde{y}}(\tilde{p}_t,\tilde{s}_t)\epsilon_{jt}^y$,
where $\mu_{\tilde{p}}(\tilde{s})$ and $\sigma_{\tilde{p}}(\tilde{s})$ are functions for the conditional mean and standard deviation for $\tilde{p}_t$, $\mu_{\tilde{y}}(\tilde{p},\tilde{s})$ and $\sigma_{\tilde{y}}(\tilde{p},\tilde{s})$ are functions for the conditional mean and standard deviation for $\tilde{y}_{jt}$, while $\epsilon_{t}^p$ and $\epsilon_{jt}^y$ are independent error terms with mean 0 and standard deviation 1. To mimic the sample size from the real corn crop dataset, we assume $n_t=500$ for $t = 1,\cdots,T$ and $T=100$ throughout the simulation study. We also perform a total of $M = 100$ Monte Carlo iterations of our simulation study. We simulate $\tilde{s}_t \iid Beta(7,44)$ for $t = 1,\cdots,T$ where $Beta(\cdot,\cdot)$ is a Beta distribution, chosen to represent the distribution of the observed $\tilde{s}_t$ from the empirical corn crop dataset.  \Copy{R3CS4}{\change{More specifically, $\tilde{s}_t \in (0,1)$, which takes the same range as the Beta distribution, because $\tilde{s}_t$ represents production-normalized carryover stocks, and the hyperparameters 7 and 44 are chosen such that the first and second moments of the Beta distribution are matched with the empirical moments of the corn stocks data as outlined in Section \ref{Empirical}.}}


 To assess the robustness of our method, we consider both linear and non-linear functional forms for $\mu_{\tilde{p}}(\tilde{s})$ and $\sigma_{\tilde{p}}(\tilde{s})$ as follows:
\begin{itemize}
\item Linear mean and standard deviation functions for the harvest price:
\[
\mu_{\tilde{p}}(\tilde{s}_t) = 0.2 - 0.4\tilde{s}_t \text{ and } \sigma_{\tilde{p}}(\tilde{s})  = 0.5 - 0.5\tilde{s}_t.
\]
\item Nonlinear mean and standard deviation functions for the harvest price:
\[
    \mu_{\tilde{p}}(\tilde{s}_t)  = -0.2 + 0.4\exp{(-2\tilde{s}_t)} \text{ and }
    \sigma_{\tilde{p}}(\tilde{s}_t) = 0.5\exp{(-2\tilde{s}_t)}.
\]
\end{itemize}

We specify the mean and standard deviation for the yield as:
\[
    \mu_{\tilde{y}}(\tilde{p},\tilde{s})  = -25 + 14.45\exp(\tilde{p}_t) + 22.18\tilde{s}_t \text{ and }
    \sigma_{\tilde{y}}(\tilde{p}_t,\tilde{s}) = 33.
\] 
The choices of these functions are meant to closely represent the realized detrended corn harvest price and county-level yields in the empirical corn dataset.

It is commonly observed that the realized detrended harvest price is typically right skewed, and the detrended crop yields are significantly left skewed \citep[][among others]{KerTolhurst,PriceYuHennessy,SwintonKing}. Therefore, we choose to model the error distributions via the skewed normal distribution. The skewed normal distribution, denoted as $X \sim \mathcal{SN}(\mu,\sigma,\alpha)$, has location and scale parameters $\mu$ and $\sigma$ along with an additional skewness parameter, $\alpha$. When $\alpha > 0$ then the distribution is right skewed and when $\alpha < 0$, then the distribution is left skewed. When $\alpha = 0$ then the distribution is the traditional normal distribution. For this simulation study, we assume that
\begin{align}
    \epsilon_t^p & \iid \mathcal{SN}\left(-\left(1 - \frac{2d_p^2}{\pi}\right)^{-1/2}d_p\sqrt{\frac{2}{\pi}},\left(1 - \frac{2d_p^2}{\pi}\right)^{-1/2},\alpha_p\right), \text{ and} \label{ptilde_errors} \\
    \epsilon_{jt}^y & \iid \mathcal{SN}\left(-\left(1 - \frac{2d_y^2}{\pi}\right)^{-1/2}d_y\sqrt{\frac{2}{\pi}},\left(1 - \frac{2d_y^2}{\pi}\right)^{-1/2},\alpha_y\right), \label{ytilde_errors}
\end{align}
for $j = 1,\cdots,n_t$, and $t = 1,\cdots,T$, where $\alpha_p = 3$, $\alpha_y = -3$, $d_p = \frac{\alpha_p}{\sqrt{1+\alpha_p^2}}$, and $d_y = \frac{\alpha_y}{\sqrt{1+\alpha_y^2}}$. These specific forms of the location and scale parameters are selected to ensure theoretical errors with zero mean and unit standard deviation. The signs of skew parameters, $\alpha_p$ and $\alpha_y$, reflect  our empirical belief that crop prices are right skewed and crop yields are left skewed. 

First, we assess the accuracy of our quantile function estimation. For each Monte Carlo sample ($m = 1,\cdots,M$), we estimate $\hat{q}_{\tau_p}$ and $\hat{q}_{\tau_y}$ for $\tau_{p}$ or $\tau_{y}$ $\in \{0.1,0.25,0.5,0.75,0.9\}$, using the methods presented in (\ref{quantreg_ptilde}) - (\ref{quantreg_ytilde}). More specifically, we calculate $\hat{q}_{\tau_p}(\tilde{s}) = \boldsymbol{B}^T(\tilde{s}) \hat{\boldsymbol{\beta}}_{\tau_p}$ and $\hat{q}_{\tau_y}(\tilde{p},\tilde{s}) = \boldsymbol{B}^T(\tilde{p},\tilde{s}) \hat{\boldsymbol{\beta}}_{\tau_y}$
for given $\tilde{p}$ and $\tilde{s}$
where $\boldsymbol{B}(\tilde{p})$ and $\boldsymbol{B}(\tilde{s})$ are constructed using $r = 3$ and $K_n = 4$ with compact support $\chi_{\tilde{p}} = [-1,1]$ and $\chi_{\tilde{s}} = [0,1]$, representing the range of plausible values for $\tilde{p}$ and $\tilde{s}$, while $\hat{\boldsymbol{\beta}}_{\tau_p}$ and $\hat{\boldsymbol{\beta}}_{\tau_y}$ are calculated using (\ref{betahat_p}) and (\ref{betahat_y}), respectively. We compare these estimated quantile functions with their true quantile functions, 
    $q_{\tau_{{p}}}(\tilde{s}) = \mu_{\tilde{p}}(\tilde{s}) + \sigma_{\tilde{p}}(\tilde{s})q_{\tau_{{p}},\epsilon_{\tilde{p}}}$ and
    $q_{\tau_{{y}}}(\tilde{p},\tilde{s}) = \mu_{\tilde{y}}(\tilde{p},\tilde{s}) + \sigma_{\tilde{y}}(\tilde{p},\tilde{s})q_{\tau_{{y}},\epsilon_{\tilde{y}}}$, where $q_{\tau_{{p}},\epsilon_{\tilde{p}}}$ is the $\tau_p^{th}-$quantile of the distribution from (\ref{ptilde_errors}) and $q_{\tau_{{y}},\epsilon_{\tilde{y}}}$ is the $\tau_y^{th}-$quantile of the distribution from (\ref{ytilde_errors}).

Figures \ref{p_lin} and \ref{p_non} present the true versus estimated $\tau_{p}^{th}-$quantile functions for $\tau_{p} \in \{0.1,0.25,0.5,0.75,0.9\}$ for linear and non-linear $\tilde{p}$, respectively. For each plot, the solid black line represents the true quantile function, the solid red line represents the median of the $M$ estimated quantile functions, and the dashed red lines represent the 0.025 and 0.975 quantiles of the $M$ estimated quantile functions. Based on the outcome of these graphs, the true quantile functions appears to directly overlap the median of the $M$ estimated quantile functions. The 0.025 and 0.975 quantile function bands fall close to the true quantile functions, suggesting our method does a good job of approximating the quantile functions of detrended prices $\tilde{p}$.


We also present the true versus estimated $\tau_{y}^{th}-$ quantile functions for $\tilde{y}$ with $\tau_{y} \in \{0.1,0.25,0.5,0.75,0.9\}$ and $\tilde{s} \in \{0.08,0.103,0.133,0.167,0.201\}$, which are the 0.1, 0.25, 0.5, 0.75, and 0.9 quantiles of the Beta distribution used to simulate $\tilde{s}_t$. Figures \ref{y_non_p_lin} and \ref{y_non_p_non} plot the quantiles for $\tilde{y}$ with linear and non-linear $\tilde{p}$, respectively. Again, the true quantile functions are almost identical with the median of the $M$ estimated quantile functions, while the 0.025 and 0.975 quantile function bands fall very close to the true quantile functions, indicating our method can estimate the quantile functions of detrended yields $\tilde{y}$ well.


\Copy{R3SC6}{Recall that the primary objectives of this paper are to estimate the conditional correlation and premium as defined in equations (\ref{condcor}) and (\ref{uncondpremium}), respectively. Both of these objectives rely on the ability to sample from the joint density function $g(\tilde{y}, \tilde{p}|\tilde{s})$. Consequently, we proceed to evaluate the performance of our proposed sampling method based on SQR in estimating the joint distribution $g(\tilde{y}, \tilde{p}|\tilde{s})$.

\change{For this evaluation, we consider $\tilde{s} \in \{0.093, 0.173, 0.281\}$, which are the 0.25, 0.5, and 0.75 quantiles of the Beta distribution used to simulate $\tilde{s}_t$.}
For each $\tilde{s}$, we simulate $R$ pairs $\{\tilde{p}_r^*, \tilde{y}_r^* \}_{r=1}^R$ \change{based} on their estimated quantile functions. Using these samples, we then estimate the 2-d density $\hat{g}(\tilde{y}, \tilde{p}|\tilde{s})$ using the \texttt{R} function \texttt{kde2d} in the \texttt{MASS} package \citep{MASS}. We compare the estimated density function $\hat{g}$ to the ``true'' density function $g$ where
\begin{align}
    g(\tilde{y},\tilde{p} | \tilde{s}) & = g(\tilde{y} | \tilde{p},\tilde{s}) g(\tilde{p} | \tilde{s}) \nonumber \\
    & = \frac{1}{\sigma_{\tilde{y}}(\tilde{p},\tilde{s})} g\left(\frac{\tilde{y} - \mu_{\tilde{y}}(\tilde{p},\tilde{s})}{\sigma_{\tilde{y}}(\tilde{p},\tilde{s})} \Big| \tilde{p},\tilde{s}\right)\frac{1}{\sigma_{\tilde{p}}(\tilde{s})} g\left(\frac{\tilde{p} - \mu_{\tilde{p}}(\tilde{s})}{\sigma_{\tilde{p}}(\tilde{s})} \Big| \tilde{s}\right) \nonumber \\
    & =  \frac{1}{\sigma_{\tilde{y}}(\tilde{p},\tilde{s})} g\left(\epsilon^y | \tilde{p},\tilde{s}\right)\frac{1}{\sigma_{\tilde{p}}(\tilde{s})} g\left(\epsilon^p | \tilde{s}\right) \label{trueg}
\end{align}
where
$g\left(\epsilon^p | \tilde{s}\right)$ and $g\left(\epsilon^y | \tilde{p},\tilde{s}\right)$ are the density functions of the distributions from (\ref{ptilde_errors}), and (\ref{ytilde_errors}), respectively. 
Figures \ref{y_non_p_lin_heat2} and \ref{y_non_p_non_heat2} illustrate the comparisons for linear and non-linear $\tilde{p}$, respectively.  The relatively small squared differences demonstrate that the estimated density obtained from our draws closely aligns with the ``true'' joint densities, especially for small levels of stocks.  \change{We also present mean integrated squared error values related to both linear and non-linear $\tilde{p}$ combined with the non-linear $\tilde{y}$ in Table \ref{tbl::MISE}}

\begin{table}[htbp!]
    \centering
    \begin{tabular}{c|cc}
    \hline
         & Linear $\tilde{p}$ & Non-linear $\tilde{p}$  \\
         \hline
        $\tilde{s} = 0.093$ & $2.774 \times 10^{-6}$ & $2.774 \times 10^{-6}$ \\
        $\tilde{s} = 0.173$ & $2.634 \times 10^{-6}$ & $2.634 \times 10^{-6}$ \\
        $\tilde{s} = 0.281$ & $4.076 \times 10^{-5}$ & $6.537 \times 10^{-5}$ \\
        \hline
    \end{tabular}
    \caption{Mean integrated squared error for the differences between the estimated and ``true'' conditional joint price-yield density for the 0.25-, 0.5-, and 0.75-quantiles of $\tilde{s}$.}
    \label{tbl::MISE}
\end{table}
}




In summary, our simulation study affirms the efficacy of our SQR with penalized B-spline method in generating sample draws that accurately estimate the true joint price and yield density. 

\Copy{R3SC5}{
 \begin{figure}[p!]
    \centering
    \includegraphics[scale=1]{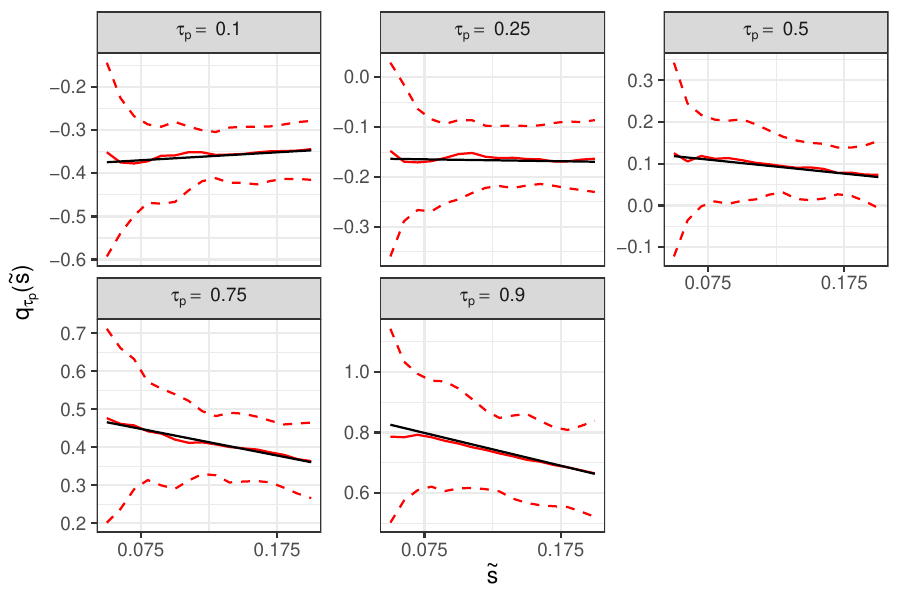}
    \caption[True and estimated quantile functions of $\tilde{p}$ from the linear $\tilde{p}$ simulation study.]{True and estimated quantile functions of $\tilde{p}$ from the linear $\tilde{p}$ simulation study. Presented are the medians (solid red lines) and 0.025 and 0.975 quantiles (dashed red lines) of the MC estimates for the quantile regression function $q_{\tau_{p}}(\tilde{s}_t)$ against the associated ``true'' quantile curve (solid black lines). Each plot represents a different $\tau_p \in \{0.1,0.25,0.5,0.75,0.9\}$.}
    \label{p_lin}
\end{figure}

\begin{figure}[h!]
    \centering
    \includegraphics[scale=1]{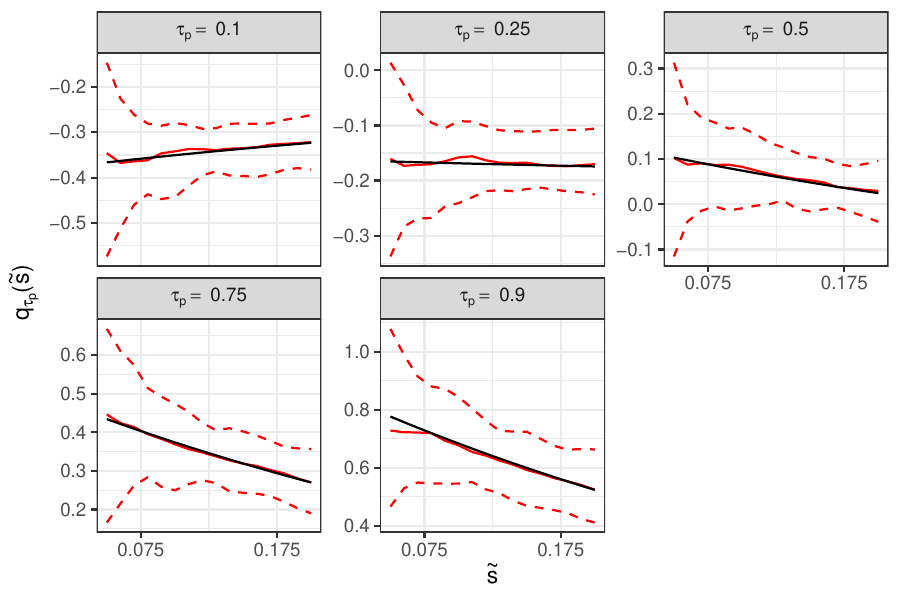}
    \caption[True and estimated quantile functions of $\tilde{p}$ from the non-linear $\tilde{p}$ simulation study.]{True and estimated quantile functions of $\tilde{p}$ from the non-linear $\tilde{p}$ simulation study. Presented are the medians (solid red lines) and 0.025 and 0.975 quantiles (dashed red lines) of the MC estimates for the quantile regression function $q_{\tau_{p}}(\tilde{s}_t)$ against the associated true quantile curve (solid black lines). Each plot represents a different $\tau_p \in \{0.1,0.25,0.5,0.75,0.9\}$.}
    \label{p_non}
\end{figure}

\begin{figure}[p!]
    \centering
    \includegraphics[scale=0.75]{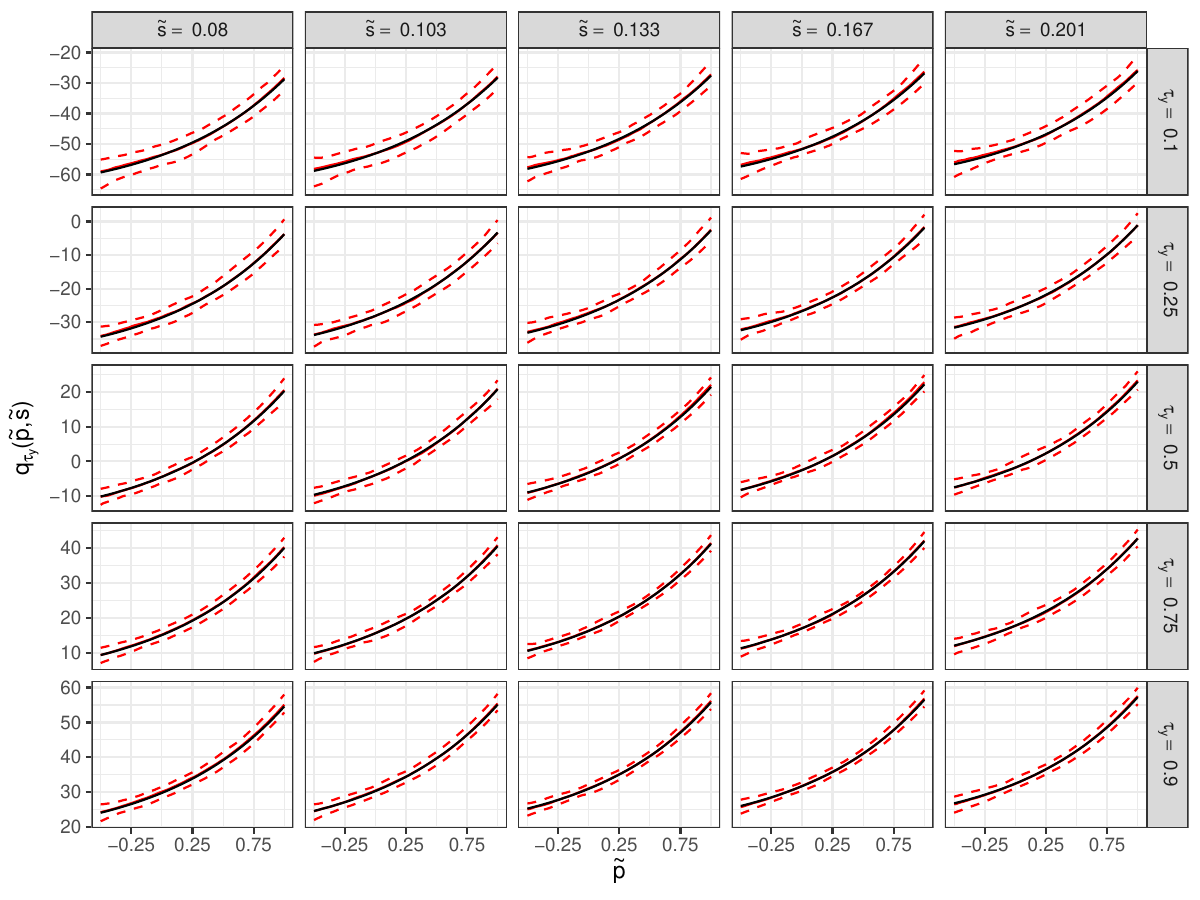}
    \caption[True and estimated quantile functions of $\tilde{y}$ from the linear $\tilde{p}$ and non-linear $\tilde{y}$ simulation study.]{True and estimated quantile functions of $\tilde{y}$ from the linear $\tilde{p}$ and non-linear $\tilde{y}$ simulation study. Presented are the medians (solid red lines) and 0.025 and 0.975 quantiles (dashed red lines) of the MC estimates for the quantile regression function $q_{\tau_{y}}(\tilde{p}_t,\tilde{s}_t)$ against the associated ``true'' quantile curve (solid black lines). Each row represents a different $\tau_y \in \{0.1,0.25,0.5,0.75,0.9\}$ and each column represents the $\{0.1,0.25,0.5,0.75,0.9\}$ quantiles for the distribution of $\tilde{s}$.}
    \label{y_non_p_lin}
\end{figure}

\begin{figure}
    \centering
    \includegraphics[scale=0.75]{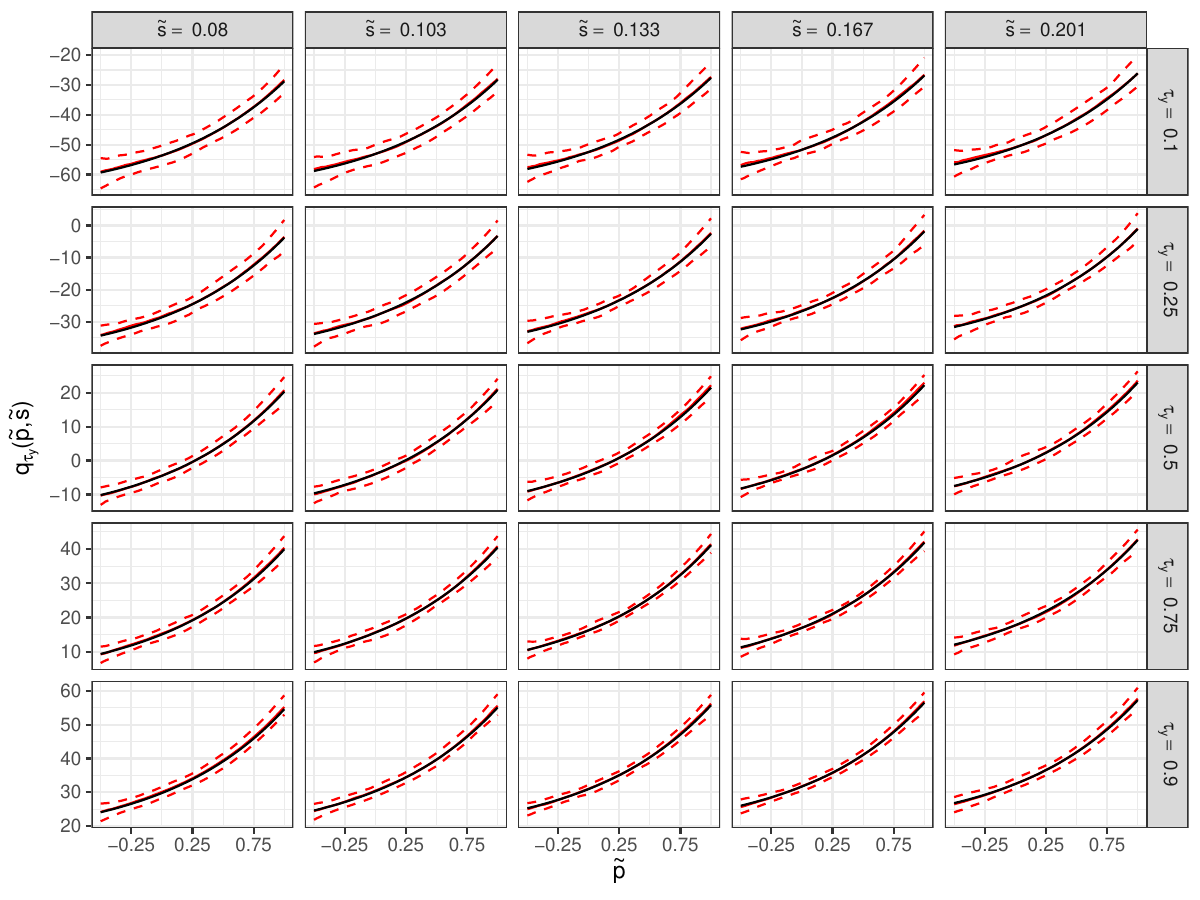}
    \caption[True and estimated quantile functions of $\tilde{y}$ from the non-linear $\tilde{p}$ and non-linear $\tilde{y}$ simulation study.]{True and estimated quantile functions of $\tilde{y}$ from the non-linear $\tilde{p}$ and linear $\tilde{y}$ simulation study. Presented are the medians (solid red lines) and 0.025 and 0.975 quantiles (dashed red lines) of the MC estimates for the quantile regression function $q_{\tau_{y}}(\tilde{p}_t,\tilde{s}_t)$ against the associated true quantile curve (solid black lines). Each row represents a different $\tau_y \in \{0.1,0.25,0.5,0.75,0.9\}$ and each column represents the $\{0.1,0.25,0.5,0.75,0.9\}$ quantiles for the distribution of $\tilde{s}$.}
    \label{y_non_p_non}
\end{figure}
}

\Copy{R3SC6a}{
\begin{figure}
    \centering
    \includegraphics{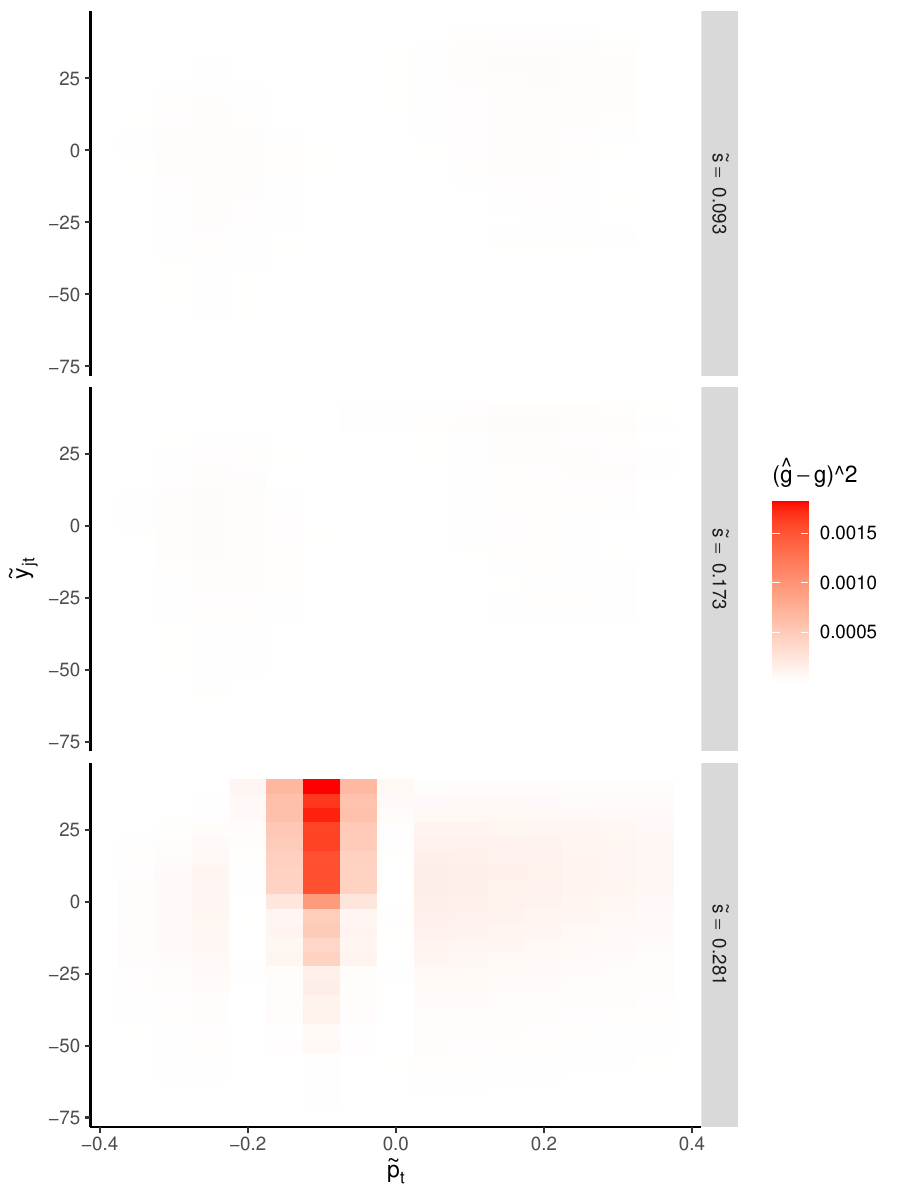}
    \caption{Heat map of the squared differences of samples from the approximate joint density $\hat{g}(\tilde{y},\tilde{p} | \tilde{s})$ compared to the ``true'' joint density $g(\tilde{y},\tilde{p} | \tilde{s})$ for linear $\tilde{p}$ and non-linear $\tilde{y}$.}
    \label{y_non_p_lin_heat2}
\end{figure}

\begin{figure}
    \centering
    \includegraphics{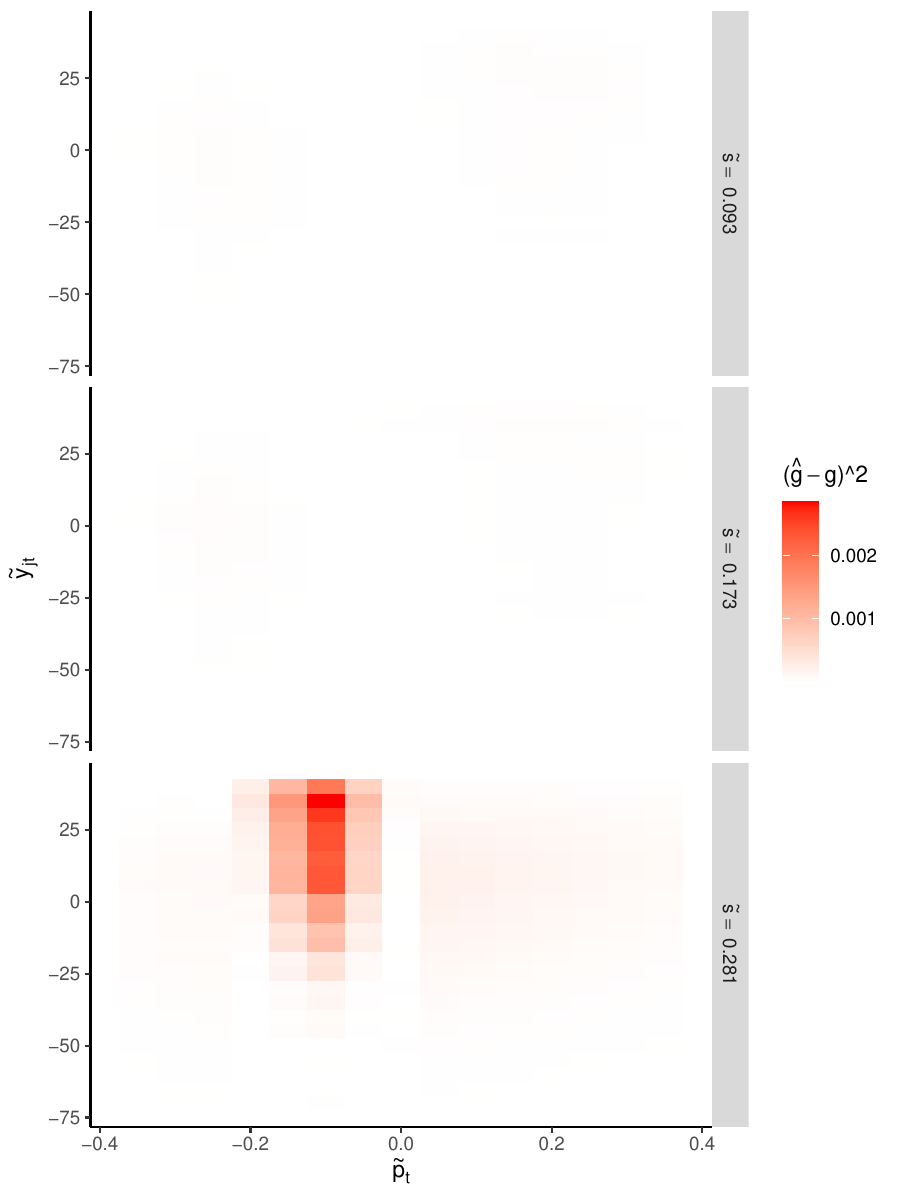}
    \caption{Heat map of the squared differences of samples from the approximate joint density $\hat{g}(\tilde{y},\tilde{p} | \tilde{s})$ compared to the ``true'' joint density $g(\tilde{y},\tilde{p} | \tilde{s})$ for non-linear $\tilde{p}$ and non-linear $\tilde{y}$.}
    \label{y_non_p_non_heat2}
\end{figure}
}

\clearpage

\section{Sampling procedure for the unconditional joint distribution} \label{unconditional}

We use a similar procedure to obtain samples from the unconditional density $g(y_{2020}, \\ p_{2020}) = g({p}_{2020})g({y}_{2020}|{p}_{2020})$. Because the univariate density for $\tilde{p}$ is not dependent on any other variables, the $\tau_p^{th}-$quantile is no longer a function and we only need to estimate the quantile scalar $q_{\tau_p}$ in addition to the quantile function $q_{\tau_y}(\tilde{p})$. In other words, for each state we have
\begin{align}
    \hat{q}_{\tau_p} & = \hat{\beta}_{\tau_p} \label{quantreg_ptilde_uncond} \text{ and}\\
    \hat{q}_{\tau_y}(\tilde{p}) & = \boldsymbol{B}^T(\tilde{p})\hat{\boldsymbol{\beta}}_{\tau_y}, \text{ where} \label{quantreg_ytilde_uncond}
\end{align}
where
\begin{align}
    \hat{{\beta}}_{\tau_p} &= \argmin_{{\beta}} \sum_{t=1}^T \rho_{\tau_p}\left(\tilde{p}_t - {{\beta}}\right) \label{betahat_p_uncond}\text{, and} \\
    \hat{\boldsymbol{\beta}}_{\tau_y} &= \argmin_{\boldsymbol{\beta}} \sum_{t=1}^T\sum_{j=1}^{n_{t}} \rho_{\tau_y}\left(\tilde{y}_{jt} - \boldsymbol{B}(\tilde{p}_t){\boldsymbol{\beta}}\right) + \frac{\lambda}{2}{\boldsymbol{\beta}}^T\boldsymbol{D}_2^T\boldsymbol{D}_2{\boldsymbol{\beta}}\label{betahat_y_uncond}.
\end{align}
Again, $\lambda$ is chosen via generalized approximate cross validation \citep{GACV}; while we could include the penalty term in (\ref{betahat_p_uncond}), no function needs to be smoothed and the value of the smoothing parameter $\lambda$ does not impact the value of our estimate $\hat{\beta}_{\tau_p}$.

By the inverse probability transformation, $\{\tilde{p}_r^* = \hat{q}_{\tau_{p,r}},\tilde{y}_r^* = \hat{q}_{\tau_{y,r}}(\tilde{p}_r^*)\}_{r=1}^R$ are $R$ independent samples from $g(\tilde{y},\tilde{p})$ where $\{\tau_{p,r}\}_{r=1}^R$ and $\{\tau_{y,r}\}_{r=1}^R$ are $Uniform(0,1)$ random variables, and using the methodology outlined in Section \ref{InflationAdj} of the online supplement, $\{y_{2020,r}^*,p_{2020,r}^*\}_{r=1}^R$ become $R$ independent samples from $g(y_{2020},p_{2020})$.

\section{Empirical Soybean Results}\label{SoybeanResults}

This section provides the empirical results for the soybean crop dataset, using the same timeline in the corn crop dataset in section \ref{Empirical} of the main text. Note, the conclusions of the soybean results are similar to those of the corn results, but noisier.

\renewcommand{\thepage}{A-\arabic{page}}
\setcounter{page}{12}

\begin{figure}[htbp!]
    \centering
    \includegraphics[scale=0.8]{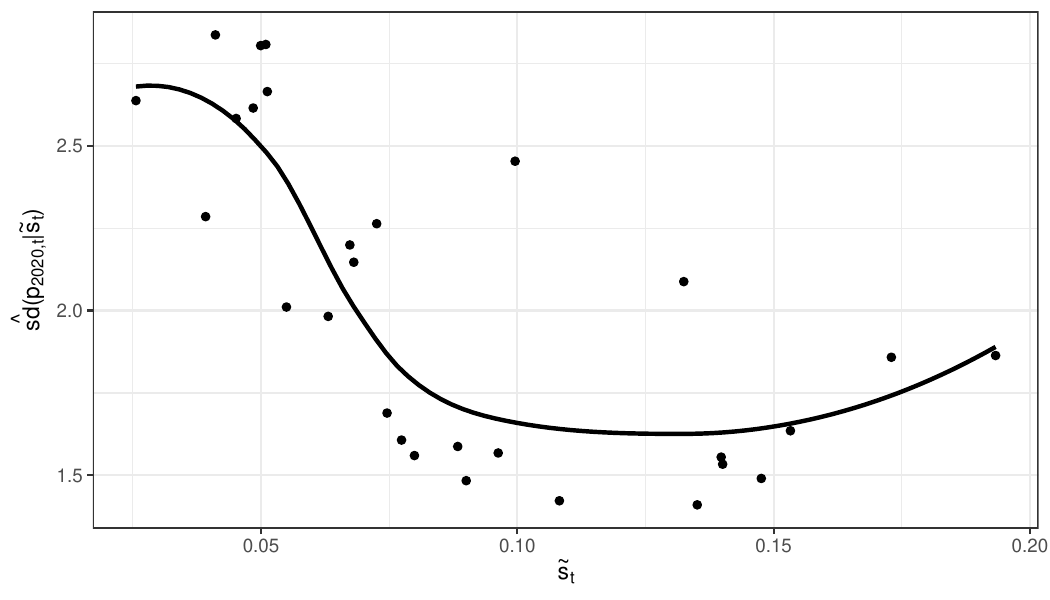}
    \caption{Standard deviation of harvest price  for soybean crops from 1990-2018 (in 2020 units). The dashed red line is the estimate of the unconditional standard deviation and the solid black line is the estimate of the conditional standard deviation function.}
    \label{Soybean_Var}
\end{figure}

\begin{figure}[htbp!]
    \centering
    \includegraphics[scale=0.8]{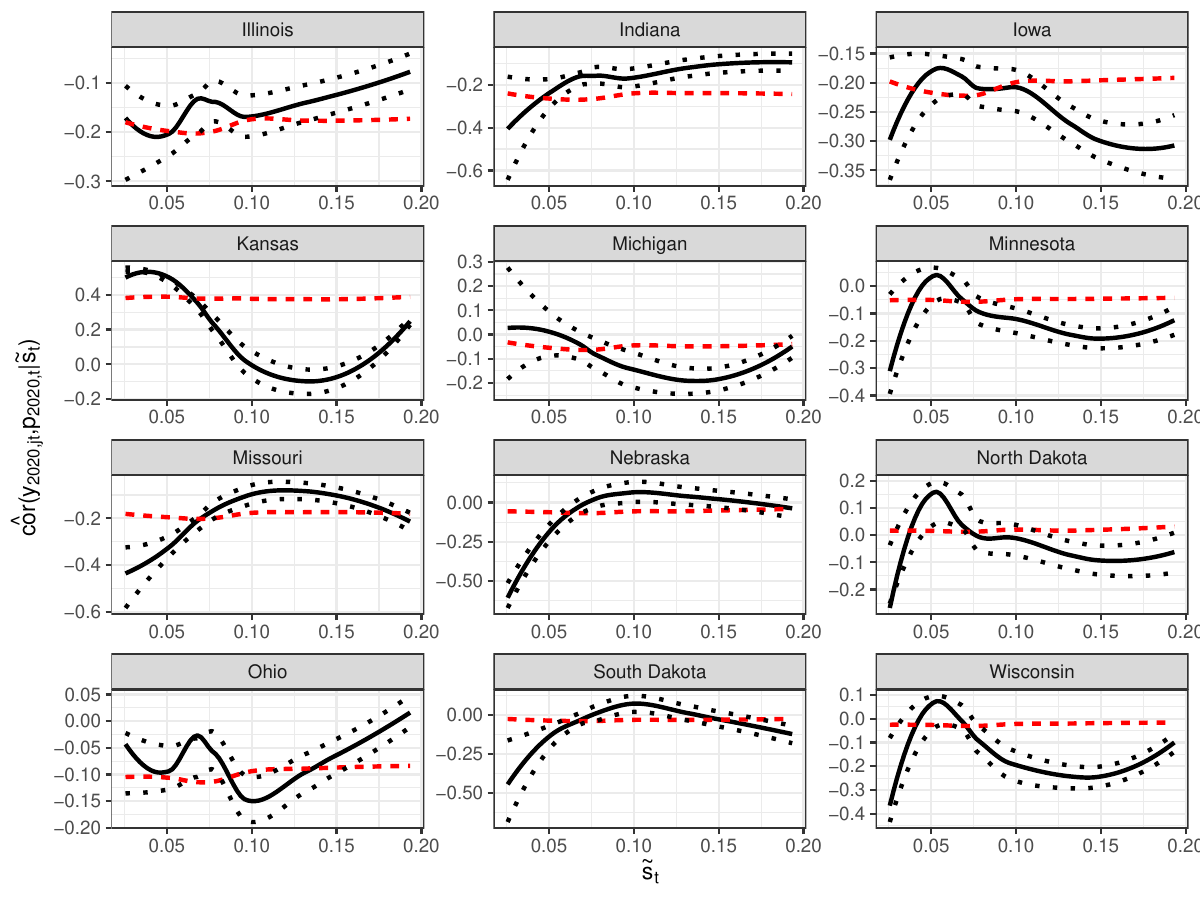}
        \caption{Correlation between harvest price and county-level yield for soybean crops by state from 1990-2018 (in 2020 units). The dashed red line is the LOESS smoothed estimate of the unconditional correlation, the solid black line is the LOESS smoothed estimate of the conditional correlation function, and the dotted black lines are the 95\% confidence bands for the conditional correlation function.}
            \label{Cor_soybean_state}
\end{figure}

\begin{figure}[htbp!]
         \centering
         \includegraphics[scale=0.9]{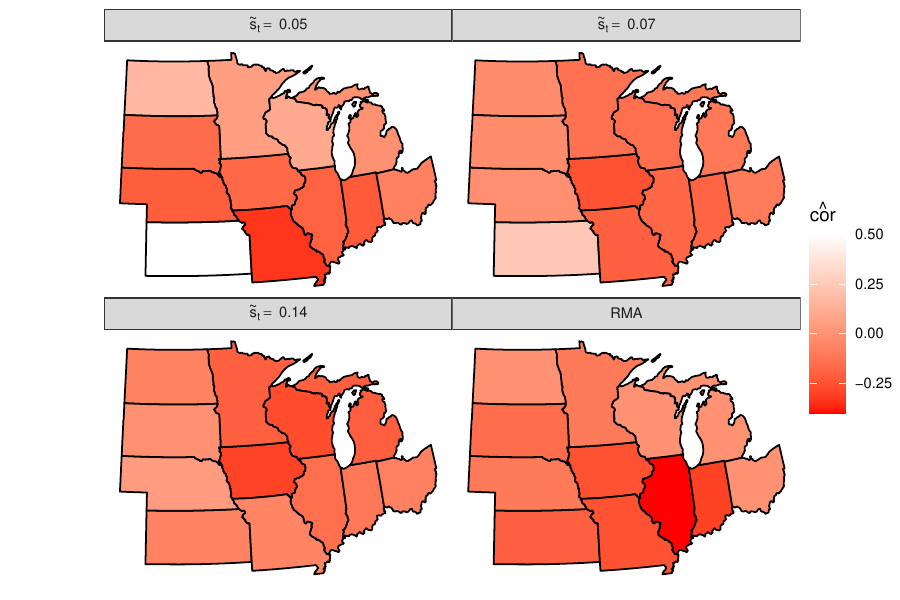}
    \caption{Heat map of the conditional correlation between harvest price and county-level yield for soybean crops across states for the observed 0.2- (left panel), 0.5-(middle panel), and 0.8- (right panel) quantiles of leftover stocks. Darker shades of red represent states with more negative correlation.}
         \label{state_map}
\end{figure}

\Copy{R1C26b}{
\begin{table}[htbp!]
\centering
\begin{tabular}{ll}
  \hline
  State & Corn \\ 
  \hline
  Illinois & 0.59 (0.56, 0.62)  \\ 
  Indiana & 0.37 (0.34, 0.41)  \\ 
  Iowa & 0.18 (0.15, 0.22) \\ 
  Kansas & 0.53 (0.50, 0.57)  \\ 
  Michigan & 0.23 (0.18, 0.28) \\ 
  Minnesota & 0.56 (0.53, 0.60)  \\ 
  Missouri & 0.24 (0.20, 0.28) \\ 
  Nebraska & 0.39 (0.35, 0.43) \\ 
  North Dakota & 0.29 (0.23, 0.35)  \\ 
  Ohio & 0.16 (0.12, 0.20) \\ 
  South Dakota & 0.45 (0.40, 0.49) \\ 
  Wisconsin & 0.40 (0.36, 0.44) \\ 
   \hline
\end{tabular}
\caption{Point estimate and 95\% confidence intervals for $\rho_{1,c_j}$ for soybean by state.}
\label{AutoCorSoybeanTable}
\end{table}}

\begin{table}[ht]
\centering
\begin{tabular}{lrr}
  \hline
State & 70\% Coverage & 85\% Coverage \\ 
  \hline
Illinois & 0.9879 & 0.2291 \\ 
  Indiana & 0.9999 & 0.5000 \\ 
  Iowa & 0.9997 & 0.9320 \\ 
  Kansas & 0.9997 & 0.9988 \\ 
  Michigan & 0.9879 & 0.6445 \\ 
  Minnesota & 0.7709 & 0.0680 \\ 
  Missouri & 0.9997 & 0.8675 \\ 
  Nebraska & 0.9693 & 0.3555 \\ 
  North Dakota & 0.9879 & 0.9320 \\ 
  Ohio & 0.9997 & 0.3555 \\ 
  South Dakota & 0.6445 & 0.5000 \\ 
  Wisconsin & 0.3555 & 0.2291 \\ 
   \hline
\end{tabular}
\caption{P-values for the premium insurance rating game for testing the three-channel premium rate method against the two-channel method for the Soybean crop dataset.}
\label{tbl::RatingGame_Soybean}
\end{table}

\end{document}